\documentclass[12pt]{iopart} 
\usepackage{cite} 
\usepackage{graphicx} 
\usepackage[colorlinks=true, linkcolor=blue, citecolor=blue]{hyperref}
\usepackage{amsmath} 
\usepackage{siunitx} 
\newcommand{\NBTBT}{Na$_{0.5}$Bi$_{0.5}$TiO$_3$-6BaTiO$_3$}

\newcommand{\EF}{E_\mathrm{F}}
\newcommand{\DEC}{^{\circ} \mathrm{C}}
\newcommand{\EVB}{E_\mathrm{VB}} 
\newcommand{\ECB}{E_\mathrm{CB}}

\begin{document}

\title[Energy gap of Ferroelectrics]{How semiconducting are ferroelectrics: The fundamental, optical and transport gaps of Na$_{0.5}$Bi$_{0.5}$TiO$_3$-BaTiO$_3$ and NaNbO$_{3}$}

\author{Pengcheng Hu$^1$, Nicole Bein$^1$, Chinmay Chandan Parhi$^2$, Tadej Rojac$^3$, Barbara Mali\v{c}$^3$, Mohammad Amirabbasi$^4$, Anton Volodin$^4$, Karsten Albe$^4$, Jurij Koruza$^2$ and Andreas Klein$^1$} 
\address{$^1$ Institute of Materials Science, Electronic Structure of Materials, Technical University of Darmstadt, Otto-Berndt-Str 3, Darmstadt 64287, Germany}
\address{$^2$ Institute for Chemistry and Technology of Materials, Graz University of Technology, Stremayrgasse 9, Graz 8010, Austria}
\address{$^3$ Electronic Ceramics Department, Jo\v{z}ef Stefan Institute, Jamova cesta 39, Ljubljana, 1000, Slovenia}
\address{$^4$ Institute of Materials Science, Materials Modelling, Technical University of Darmstadt, Otto-Berndt-Str. 3, Darmstadt 64287, Germany}
\ead{aklein@esm.tu-darmstadt.de}

\begin{abstract}
The energy gap is a fundamental property of materials, directly related to their optical and electronic properties. The energy gap of ferroelectric compounds and its adjustment by compositional variation has particularly attracted attention in recent years due to potential application in energy conversion and/or catalytic devices. It is demonstrated that it is necessary to distinguish between the fundamental gap, $E_{\rm g}^{0}$, the optical gap, $E_{\rm g}^{\rm opt}$, and the transport gap, $E_{\rm g}^{\rm tr}$, of ferroelectrics, which can differ significantly. The situation is comparable to those in organic semiconductors and emerges from the presence of localized charges. The fundamental gap is a ground state property, i.e.\ the energy difference between the maximum of the fully occupied valence band and the minimum of the completely empty conduction band. In contrast, the optical and transport gaps are excited state properties involving localized (polaronic) electrons and/or holes at energies considerably different from the band edges. This work illustrates how the different energy gaps of ferroelectrics can be determined by combining optical measurements, X-ray photoelectron spectroscopy and temperature and oxygen partial pressure dependent electrical conductivity measurements. We determine fundamental gaps of $\approx 4.5\,$eV for both materials, optical gaps of $3.25-3.45\,$eV/$3.5\,$eV and electrical gaps of $\approx 1.4\,$eV/$3.3\,$eV for Na$_{0.5}$Bi$_{0.5}$TiO$_3$-BaTiO$_3$/NaNbO$_{3}$, respectively. 
\end{abstract}


\submitto{\RPP}

\maketitle

\section{Introduction}

\subsection{Semiconducting ferroelectrics}

\begin{figure}[ht]
    \centering
    \includegraphics[width=0.7\linewidth]{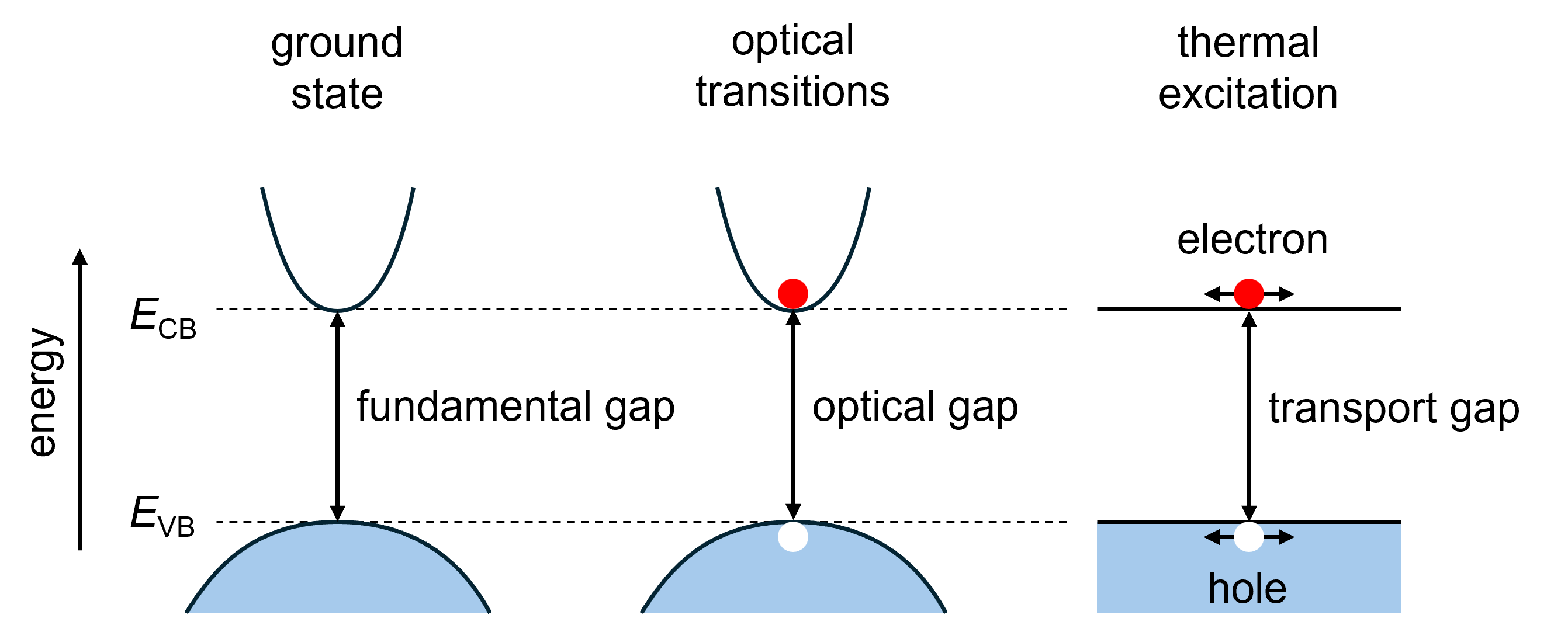}
    \caption{The energy gaps in classical semiconductors. The fundamental gap separates the occupied states in the valence band from the unoccupied states in the conduction band. The energies of the states do not change when the states are occupied by electrons or holes. The fundamental gap is therefore also relevant for optical transitions and the thermal generation of mobile charge carriers.  }
    \label{semiconductor}
\end{figure}

The opto-electronic properties of materials are intimately connected to their electronic structure \cite{sze,Yu01}. The energy gap, which separates occupied valence bands from unoccupied conduction bands, is crucial for opto-electronic applications such as photo\-de\-tec\-tors, solar cells, light-emitting diodes, semiconductor lasers and photocatalysts. All these applications are enabled by semiconducting materials, in which electrical and optical properties are governed by the electronic structure of the valence and conduction band and their occupation by electronic charge carriers. In the ground state, occupied states at the top of the valence band are separated from empty states at the minimum of the conduction band by the fundamental energy gap. In semiconductors, the occupation of the valence and conduction band by holes and electrons is conventionally described by the occupation of electronic states, which do not change their energy upon occupation of the state. In this case, the optical and transport gaps, which are related to optical and thermal transitions of electronic carriers, respectively, are identical to the fundamental gap as illustrated in Fig.~\ref{semiconductor}. 

\begin{figure}[ht]
    \centering
    \includegraphics[width=10.5cm]{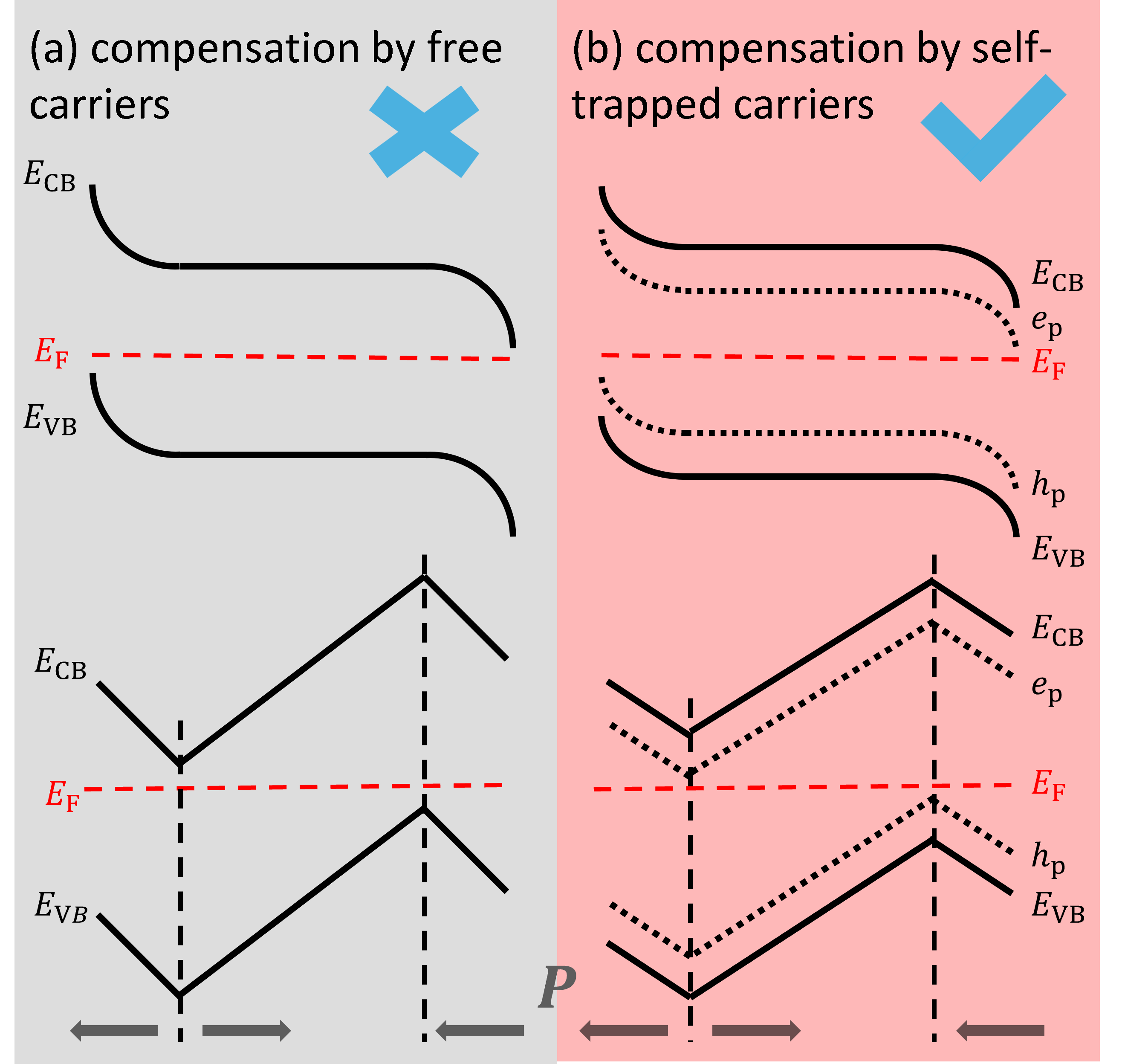}
    \caption{Energy band diagrams for a ferroelectric material with polarization (P) and charged domain walls, as proposed in \cite{li2014photocatalysts} and \cite{khan2016ferroelectric}, where the polarization is compensated by free electrons and holes and the Fermi level varies through the complete energy gap (a); (b) Hypothetical energy band diagram with upper and lower Fermi energy limits determined by self-trapped electrons and holes, where the polarization is compensated by trapped charges and the Fermi level varies only between the electron's and the hole's polaron levels $e_{\mathrm{p}}$ and $h_{\mathrm{p}}$. }
    \label{screening}
\end{figure}

In ferroelectric materials, which exhibit spontaneous polarization switchable in symmetry-determined crystallographic directions, (free) electrons and holes in the conduction and valence bands have been invoked to discuss conductive domain walls or photocatalytic effects \cite{sluka12,zuo14,Li14} (see Fig. \ref{screening}(a)). In both cases, (free) electrons and holes are employed to compensate for the bound polarization charges of $\approx 40\,\mu {\rm C/cm^2}$, which terminate ferroelectric domains. However,  Fridkin already stated that the concentration of free electrons and holes in ferroelectrics is typically negligible as the electronic charge carriers are trapped \cite{fridkin}. In more ionically bonded solids, such as ferroelectric oxides, ionic defects and trapped electrons and holes need to be taken into account as well \cite{Klein23}; these are also referred to as polarons. Trapped electronic charge carriers are also discussed for charge transport in ferroelectrics \cite{bidault1995polaronic, robertson95, akkopru21, schirmer09, schirmer11}, but their role in the energy band diagram and their connection to the energy gap are hardly considered. Figure~\ref{screening} illustrates some anticipated consequences of trapped charge carriers on the polarization screening by mobile electronic carriers. There are two major differences between free and trapped charge carriers:  i) the concentration of free electrons as determined by the occupation of the electronic states near the edges is connected to the effective density of states in the bands \cite{sze}:
\begin{equation}
    N_{V,C} = 2 \left( \frac{2 \pi m_{\rm h,e}^\ast k_{\rm B}T}{h^2}\right) ^{3/2}, 
\end{equation}
where $k_{\rm B}$, $T$, and $h$ are Boltzmann's constant, temperature, and Planck's constant, and $m_{\rm e,h}$ correspond to the effective masses of holes and electrons, respectively. For typical values of the effective masses of $1-5$, the effective density of states is $\approx 10^{19}\,{\rm cm^{-3}}$. This value is orders of magnitude lower than the density of polaron sites, which is equal to the atom density of the trapping species ($\approx 10^{22}\,{\rm cm^{-3}}$). Due to the low density of states of free electrons/holes, the width of the region required to provide the charges for screening polarisation ($\approx 40\,\mu {\rm C/cm^2}$) is several $100\,$nm  \cite{sluka12,zuo14}. In contrast, a unit cell thick layer is sufficient in the case of trapped charges; ii) The variation of the Fermi level ranges from the valence band maximum to the conduction band minimum in the case of free charges. For trapped charges, the Fermi level can only vary between the electron's and the hole's polaron levels, which will limit, e.g.\ the photovoltages developing upon illumination. The differences in polarisation screening by free/trapped charges are illustrated by energy band diagrams in Fig.~\ref{screening}.

The effect of trapped electronic charges on transport properties is straightforward, but their role on optical transitions is less clear and likely also depends on the material. The underlying effects are particular crucial for optoelectronic and photocatalytic applications of ferroelectrics \cite{spanier16, bai18, choi09, lee11, kreisel12, grinberg13, ji13}. A major issue of ferroelectrics is their low sensitivity to visible light, as their ``energy gaps'', determined by optical techniques, are typically $> 3\,$eV. Tuning the energy gap hence became an important issue \cite{grinberg13,das18}. Therefore, it is inevitable to understand the nature of the band gap, particularly what is measured by optical techniques, and how this gap relates to the energy levels involved in electrical transport and the fundamental gap from standard density functional theory calculations.

\subsection{The nature of the energy gap}

Traditionally, the energy gap is measured by optical techniques such as transmission, reflectance and/or ellipsometry \cite{fox2010optical}. For non-transparent polycrystalline ferroelectrics, diffuse reflectance with an application of the Kubelka-Munk evaluation and Tauc plots is often the method of choice \cite{kubelka1931beitrag,tauc1966optical}. This technique employs an optical transition and inducing the formation of an electron-hole pair. The resulting excited state has to be distinguished from the ground state, which corresponds to a fully occupied valence band and a completely empty conduction band. The term \emph{fundamental energy gap} is used here for the energy difference between valence band maximum, $\EVB$, and conduction band minimum, $\ECB$, in the ground state. The fundamental gap related to the ground state of a material is denoted in this manuscript as $E_{\rm g}^0$. 

An important modification of the electronic states from their ground state is induced by electron-phonon interactions, which induce local lattice distortions \cite{Austin69, franchini21, Stone07, Shluger93, natanzon20} and result in trapping of the charge carriers. In conventional semiconductors such as Si and GaAs, the effect of lattice distortion is typically neglected as the mobility of charge carriers can be described by classical movement inside the energy bands. In case of charge trapping, occupation of a trap occurs once the Fermi level crosses a certain level, which is denominated as charge transition level of the traps, $E_{\rm T}^{\rm e}$ for electrons and as $E^{\rm h}_{\rm T}$ for holes. These charge transition levels constitute upper and lower limits of the Fermi level, $\EF$ \cite{Klein23}. The energy difference between the charge transition levels of electron and hole traps hence defines the \emph{transport gap} of the material: $E_{\rm g}^{\rm tr} = E_{\rm T}^{\rm e} - E_{\rm T}^{\rm h}$. The transport gap is directly relevant for photo-electric and photo-catalytic applications. For the former, the splitting of the quasi Fermi levels, which determines the photovoltage of a device \cite{wuerfel}, is limited by the transport gap. For photocatalytic processes, the charge transfer at the active interface occurs from the trap levels of the solid and not from the band edge energies \cite{Klein23b}. The latter explains the poor efficiency of Fe$_2$O$_3$ for solar water splitting despite an apparently ideal energy gap of $2.2\,$eV, where electron trapping limits the Fermi level to values of $\ECB - \EF > 0.5\,$eV  \cite{lohaus18}. 

The transport is determined by non-interacting trapped electrons and holes and has thus to be distinguished from the energy gap measured by optical transitions, the \emph{optical energy gap}, $E_{\rm g}^{\rm opt}$, where Coulomb attraction between electrons and holes can result in the formation of bound electron-hole pairs, so-called excitons. The exciton binding energies in conventional semiconductors are typically a few meV and therefore only noticeable at cryogenic temperatures, where they show up as narrow absorption peaks at photon energies slightly below the fundamental gap \cite{Yu01}. ZnO exhibits a slightly higher exciton binding energy of $59\,$meV, which can be observed at room temperature \cite{thomas60}. Excitonic absorption features are also pronounced in layered transition metal di-chalcogenides \cite{wilson69}. 

\begin{figure}[htp]
    \centering
    \includegraphics[width=0.7\linewidth]{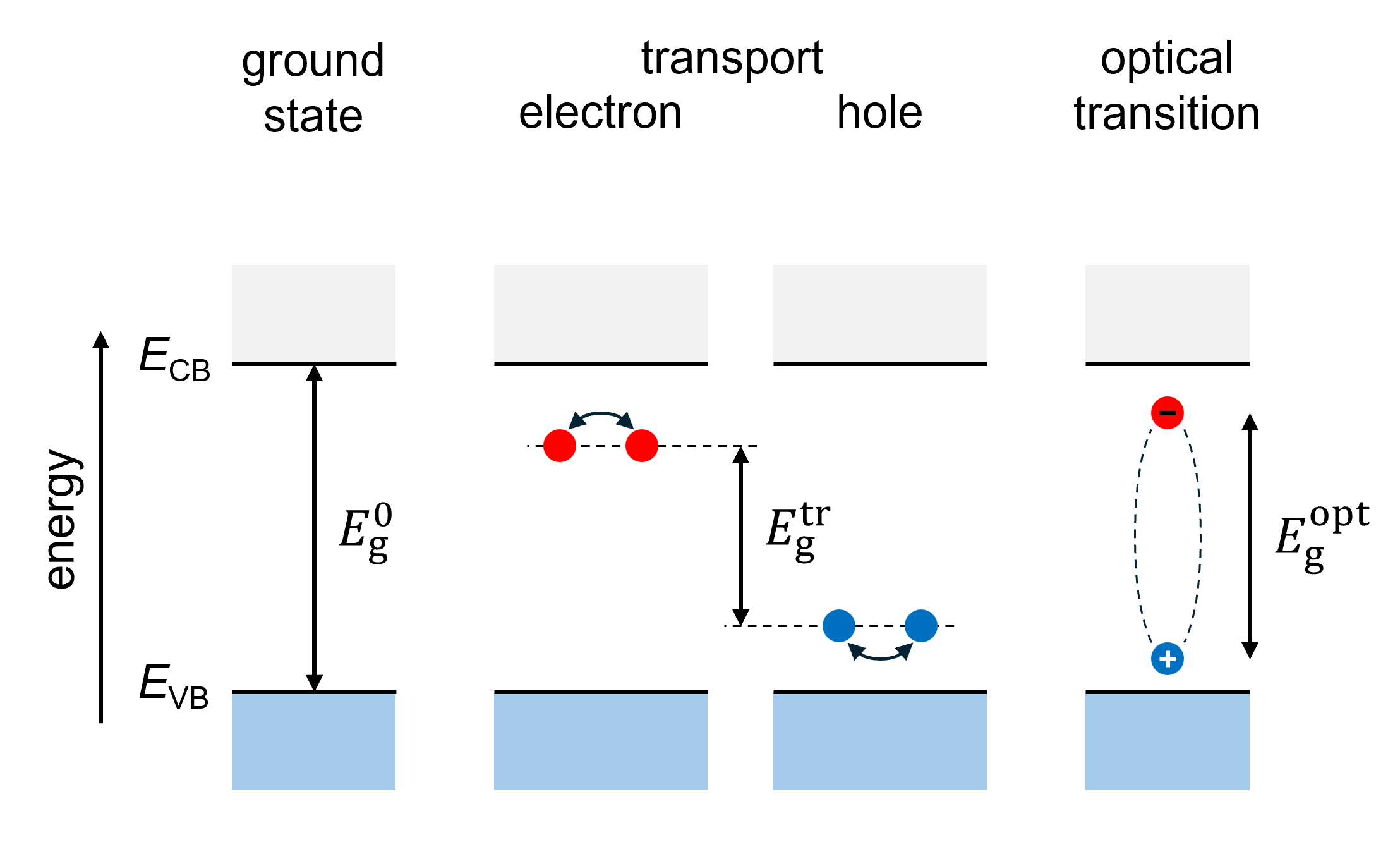}
    \caption{Schematic comparison of the fundamental gap ($E_{\rm g}^{0}$), corresponding to the ground state energy difference of the valence band maximum and the conduction band minimum, the transport ($E_{\rm g}^{\rm tr}$) emerging from charge trapping and the optical gap ($E_{\rm g}^{\rm opt}$), related to exciton formation, of ferroelectrics. The curved solid lines with arrows indicate hopping transport while the curved dashed lines indicate the Coulomb attraction of the photo-excited electron-hole pair forming an exciton. The order of the magnitudes of the different energy gaps, $E_{\rm g}^{0} > E_{\rm g}^{\rm opt} > E_{\rm g}^{\rm tr}$, follows the experimental observations described in this work. }
    \label{gaps}
\end{figure}

Due to the high permittivity of ferroelectrics, there is a strong tendency for the localisation of charge carriers and the formation of polarons in ferroelectrics \cite{franchini21}, since the large dielectric response of polar oxides reflects strong coupling between electronic charges and lattice polarization. This coupling allows local lattice distortions to effectively screen the carrier’s Coulomb potential, lowering its self-energy and stabilizing a localized state \cite{Shluger93, Stone07}. Examples include hole polarons at oxygen sites in BaTiO$_3$ \cite{erhart14, Traiwatt18, schirmer11}, LiNbO$_3$ \cite{schirmer11, schirmer09, schmidt08} and KTaO$_3$ \cite{bidault95}, hole polarons at Pb-sites in Pb(Zr,Ti)O$_3$ \cite{robertson95}, electron polarons on Ti sites in Pb(Zr,Ti)O$_3$ \cite{robertson95, ghorbani22} and Nb-sites in LiNbO$_3$ \cite{schirmer09}. Although the existence of polarons has been reported, the experimental determination of trapping energies and transport gaps is largely missing and often misinterpreted using the activation energies determined from, e.g., temperature dependent electric conductivity. Also exciton formation is known for ferroelectrics and considerable exciton binding energies of $\approx 1\,$eV were reported in literature for LiNbO$_3$ \cite{schmidt08}, KNbO$_3$ \cite{schmidt19} and NaNbO$_3$ \cite{bein22}. A tentative scheme for the three different energy gaps in ferroelectrics is depicted in Fig. \ref{gaps}.  

The distinction between the fundamental, transport and optical energy gaps is well-known for organic semiconductors, where excess charges are localized on small molecules or molecular units of larger molecules (polymers) \cite{heeger88,bruetting12}. The occupation of non- or anti-bonding orbitals (lowest unoccupied molecular orbital: LUMO) by an electron or the removal of an electron from the highest occupied molecular orbital (HOMO) results in a strong electronic polarization of the molecule and in a distortion of the chemical bonds associated with a structural relaxation. Trapping energies in organic molecules are naturally high with magnitudes up to few hundred meV. For the same reason, exciton binding energies in organic semiconductors are up to $1.5 \,$eV depending on the size of the molecules \cite{bruetting12,knupfer03}. In organic semiconductors, the order of the energy gaps is $E_{\rm g}^{0} > E_{\rm g}^{\rm tr} > E_{\rm g}^{\rm opt}$, which is different from the order anticipated in Fig.~\ref{gaps}.

\subsection{Ingredients for the determination of the energy gaps}

While the determination of the optical gap by means of optical or electron spectroscopies  for ferroelectrics is the same as for all materials, this work will demonstrate how the fundamental and transport gaps of ferroelectrics can be determined experimentally by a combination of X-ray photoelectron spectroscopy (XPS) and electrical transport measurements. The determination of the gaps is exemplified for two compounds exhibiting polaronic electronic conduction of both electrons and holes: NaNbO$_3$ and 0.94(Na$_{0.5}$Bi$_{0.5})$TiO$_3$-0.06BaTiO$_3$\ (NBT-6BT), where 6\% is mole fraction of BaTiO$_3$. Key ingredients enabling the determination of the fundamental and transport gap are in both cases i) the direct determination of the charge transition level of either the electron or the hole polaron by means of XPS measurements; ii) the derivation of the position of the valence or conduction band edge relative to the trap level from the activation energy of electronic conductivity in an extrinsic temperature regime with fixed polaron contributions; iii) the derivation of the Fermi level position in a temperature regime with temperature dependent polaron concentration from the activation energy of electronic conductivity; iv) the determination of the trapping energy of the other polaron parameters from the activation energies of electronic conduction of a sample exhibiting a transition from n- to p-type behavior at a certain temperature. 

The approach requires samples of the same material with different types of electronic conduction, i.e.\ effectively donor-doped and effectively acceptor-doped materials. Acceptor-doping often results in the formation of oxygen vacancies and the occurrence of oxygen ion conduction. This is particularly the case for NBT-based materials, where oxygen ion conduction dominates over (p-type) electronic conduction in effectively acceptor-doped compositions \cite{li2014family}. As the determination of the energy gaps requires the analysis of the electronic conduction, it is mandatory to discriminate between electronic and ionic conduction. In this work, the separation is performed by comparing alternating current (AC) and direct current (DC) measurements. AC measurement reveal the total conductivity (electronic + ionic), whereas dc measurements are selectively probing electronic conductivity due to the oxygen blocking nature of the metal electrodes (such as gold, silver and platinum) \cite{baiatu1990dc}. The separation between n- and p-type electronic conduction in (Na$_{0.5}$Bi$_{0.5})$TiO$_3$-based compositions is achieved by variation of the sample composition as demonstrated by Sinclair and coworkers \cite{li2014family,yang2018defect,li2016controlling}. For NaNbO$_3$, the variation is accomplished by a comparison between nominally undoped and donor- (Ca-) doped samples. 

The following of the article is organized as follows: Section \ref{methods} provides details of sample preparation and characterization and of density functional theory calculations, which are performed to compare the fundamental band gap with those derived from experiments. The model used to extract the trap levels from the electrical conductivity measurements is outlined in section~\ref{polaron}. Results obtained from optical, XPS, electrical measurements are described in sections \ref{optical}, \ref{XPS}, and \ref{electrical}, respectively. The extraction of polaron binding and formation energies is treated in section \ref{polaronenergy}, eventually resulting in the experimental determination of the transport and fundamental gaps in section \ref{TFgap} and a confirmation of the fundamental gaps by density functional theory calculations in section \ref{DFT}. Finally, section \ref{semifer} revisits the semiconducting nature of ferroelectric materials in the context of the presented results and sketches some consequences for material properties and other ferroelectric compounds.  

\section{Methods} 
\label{methods}

\subsection{Sample preparation}

\textbf{\NBTBT\ (NBT-6BT)}: 0.94(Na$_{0.5}$Bi$_{0.5}$)TiO$_3$-0.06BaTiO$_3$ (NBT-6BT) and 0.94(Na$_{0.51}$Bi$_{0.49}$)TiO$_3$-0.06BT were prepared by solid-state synthesis. These samples will be marked as 50/50 and 51/49. Powders with given purities were utilized: Na$_2$CO$_3$ (99.95\%), BaCO$_3$ (99.95\%), Bi$_2$O$_3$ (99.975\%) (all from Alfa Aesar GmbH \& Co. KG, Germany) and TiO$_2$ (99.8\%, Sigma Aldrich). After weighing the precursors as per the stoichiometry, the powders were milled with yttria-stabilized zirconia balls in ethanol for 4\,h at 250\,rpm then dried and homogenized, and finally calcined at 900$^\circ$C for 3\,h using a heating rate of 5$\,^\circ$C/min. Milling was subsequently repeated under the same conditions and was followed by one more drying step. Disk samples of 13\,mm in diameter were pressed at a uniaxial pressure of 15.4\,MPa, followed by an isostatic pressure of 300\,MPa. The samples were placed in a closed alumina crucible with sacrificial powder and were sintered at 1150$\,^\circ$C for 3\,h using a ramp rate of 5$\,^\circ$C/min. 

\textbf{NaNbO$_3$}: The polycrystalline undoped NaNbO$_3$ (NN) and Ca-doped NaNbO$_3$ (CaNN) samples were also prepared by solid-state reaction. Na$_2$CO$_3$ (99.9\%, ChemPur) and orthorhombic Nb$_2$O$_5$ (99.9\%, Sigma-Aldrich) powder were used as starting materials, with calcium carbonate (CaCO$_3$, 99.95\%, Alfa Aesar) added for doping. Before mixing, the powders were milled first by attrition milling with isopropanol and yttria-stabilized zirconia balls for 2\,h at 500 rpm in case of Nb$_2$O$_5$ and by planetary ball milling with acetone and yttria-stabilized zirconia balls for 4h at 200 rpm for Na$_2$CO$_3$ and CaCO$_3$. Then the powders were dried at 100$\,\DEC$ for 1\,h and at 200$\,\DEC$ for 2\,h, sieved, and dried again at 100$\,\DEC$ for 2\,h. The dried powders were weighed in a Labmaster 130 glove box (MBraun, Garching, Germany), mixed, and homogenized for 4\,h. In case of the doped samples, Na$_{0.99}$Ca$_{0.01}$NbO$_3$ (CaNN), CaCO$_3$ was added according to a 1\,mol\% doping. The homogenization and all following milling steps were done with acetone and yttria-stabilized zirconia balls via planetary ball milling followed by drying at 105$\,\DEC$ for 1\,h and at 200$\,\DEC$ for 2\,h, sieving, and drying at 200$\,\DEC$ for 2\,h. After homogenization, the powders were uniaxially pressed to pellets and calcined for 4\,h at 700$\DEC$ in case of NaNbO$_3$ in an alumina crucible. For Na$_{0.99}$Ca$_{0.01}$NbO$_3$, a calcination temperature of 950$\,\DEC$ was chosen. A second calcination step was performed at the same conditions after crushing the samples and milling them again for 3h. After the final milling process for 3\,h, pellets with a diameter of 8\,mm were pressed isostatically with 200\,MPa and sintered in air for 2\,h at $1250 \DEC$ in case of pure and at $1320\,\DEC$ for Ca-doped NaNbO$_3$.

%X-ray diffraction (XRD) was performed on powder samples, prepared by crushing and annealing the sintered pellets. A Rigaku-600 X-ray diffractometer and a Cu-source (Cu K$\alpha$1 $\lambda$ = 1.5409 $\AA$) is used in transmission geometry.

\subsection{Electrical conductivity measurements}

\textbf{NBT-6BT}: The ceramic pellets were initially ground with sandpaper to achieve a thickness of $0.30 - 0.35\,$mm and then cut into small square-shaped pieces (length = $3.0\,$mm). To reduce the residual stress from grinding, the pellets were annealed at 450$\,^\circ$C for 1 h, with heating and cooling rates of 5$\,^\circ$C/min in a box furnace. Platinum (Pt) electrodes were then sputtered onto both sides of each pellet using a Quorum Q300T D sputter coater (Quorum Technologies Ltd., UK). The Pt electrodes were round shaped with a diameter of 2.5\,mm for all samples.

The dynamic temperature dependence of conductivity were performed by direct current (DC) method in different atmospheres (dry air and N$_{2}$ with oxygen partial pressures of $10^5$ and $10^0$, respectively): the sample was ramped up at a constant rate of $2.5\,^\circ$C/min to $450\,^\circ$C, held at 450$\,^\circ$C for one hour, and then cooled back to room temperature at the same rate. During this thermal cycling, a fixed dc voltage of $0.5\,$V using a Keithley 6487 picoammeter (Tektronix, Inc., USA) was continuously applied, and the current through the sample was recorded. This thermal cycling was performed twice to check reproducibility. The oxygen partial pressure was monitored using the SGM5-EL electrolysis device (ZIROX Sensoren $\&$ Elektronik GmbH, Germany).

\textbf{NaNbO$_3$}: The densified samples were ground to a thickness of 0.7\,mm. Platinum electrodes were sputtered on the surfaces. The dynamic temperature dependence of conductivity were also performed by direct current (dc) method in different atmospheres (dry air and N$_{2}$ with oxygen partial pressures of $10^5$ and $10^0$, respectively): the sample was ramped up at a constant rate of $1.0\,^\circ$C/min to $500\,^\circ$C or $600\,^\circ$C, held for 2\,mins, and then cooled back to room temperature at the same rate. During this thermal cycling, a fixed dc voltage of $0.1\,$V or $1.0\,$V was applied.

\subsection{X-ray photoelectron spectroscopy} 

\textbf{NBT-6BT}:
In situ X-ray photoelectron spectroscopy (XPS): the sintered ceramic samples were first ground with sandpaper (\#800, \#1200) to 0.45\,mm thickness and then cut into a rectangular shape of 3$\times$4 mm$^2$. The samples were subsequently annealed in air at 450$^\circ$C for 1\,h to relieve the residual stress introduced by machining. The bottom electrodes of Pt with a thickness of 50\,nm were deposited with a sputter coater (Quorum Q300T D, Quorum Technologies Ltd., UK). Before the deposition of the Indium tin oxide (ITO) top electrode, a surface cleaning was carried out inside the deposition chamber of the Darmstadt Integrated System for Materials Research (DAISY-MAT) by heating in oxygen (0.5\,Pa, 400$^\circ$C, 0.5\,h) to remove adventitious carbon species. Top electrodes of 10\% Sn-doped In$_2$O$_3$ with a thickness of 2\,nm were subsequently deposited by radio frequency magnetron sputtering either at room temperature for subsequent annealing or at 400$^\circ$C. Finally, the samples were mounted onto stainless-steel sample holders allowing for separate electrical contacts to the bottom and top electrode. The ITO electrode was connected to the ground, ensuring that the Fermi energy at the top of the sample is aligned with that of the spectrometer, which serves as a binding energy reference for the spectra. XPS measurements were then executed by a Physical Electronics PHI 5700 spectrometer system (Chanhassen, MN) with monochromated Al K$\alpha$ radiation. Binding energies were calibrated using a sputter cleaned Ag foil. XPS measurements were performed either in the course of heating the samples inside the XPS chamber or by applying a positive voltage to the Pt electrode at elevated temperatures.

\textbf{NaNbO$_3$}:
For the interface experiment, the clean surfaces were first characterized in the analysis chamber using XPS. The electrode material was then deposited stepwise with increasing thickness until the substrate signals were attenuated, with XPS measurements performed after each deposition step. ITO was deposited via radio frequency (RF) magnetron sputtering at a power of 25\,W under an Ar flow of 10 sccm and a pressure of 0.5\,Pa at 400$\DEC$, resulting in a sputter rate of 5\,nm/min and a polycrystalline film. For RuO$_2$ deposition, direct current (dc) magnetron sputtering at room temperature was used to grow amorphous RuO$_2$ on the sample surface. The sputtering conditions were 10 W power and 1\,Pa pressure in an Ar/O$_2$ mixture (9.25\,sccm Ar and 0.75\,sccm O$_2$), yielding a sputter rate of 3\,nm/min. XPS analysis was performed using a Physical Electronics PHI 5700 spectrometer (Physical Electronics, Chanhassen, MN) with monochromatic Al K$\alpha$ excitation. All spectra were calibrated using a sputter-cleaned Ag foil, with the Fermi edge set to 0\,eV.

\subsection{Optical measurements}

\textbf{NBT-6BT}:
The optical band gap ($E_{\textsubscript{g}}$) of NBT-6BT was determined from diffuse reflectance spectra of powder samples (DRS) recorded with a Shimadzu UV-2600i over the wavelength range of 220–1400 nm.

\textbf{NaNbO$_3$}:
Low-loss electron-energy loss measurements (EELS) spectra were acquired using the Zeiss SESAM microscope (ZEISS GmbH, Wetzlar, Germany) at an acceleration voltage of 200\,kV in TEM mode. The microscope is equipped with an electrostatic $\Omega$-type monochromator (CEOS GmbH, Heidelberg, Germany) and the in-column MANDOLINE energy filter. EELS data were acquired with an energy resolution of 60\,meV as determined from the full width at half-maximum of the zero-loss peak (ZLP). 

\subsection{Density functional theory calculations}
\textbf{NBT}: 

All calculations were performed with the Vienna Ab initio Simulation Package (VASP) \cite{kresse1993ab,kresse1994ab,kresse1996efficiency,kresse1996efficient} using the projector-augmented-wave (PAW) method \cite{blochl1994projector,kresse1999ultrasoft}. Band gaps of NBT were computed within the GW approximation \cite{hedin1965new}. For Brillouin-zone sampling we used Monkhorst-Pack k-point meshes with the density of approximately 27\,\AA\ and 22\,\AA\ for 100- and 111-ordered structures respectively. The primitive cells for 100- and 111-ordered structures %had volumes of approximately 231 \AA$^3$ and 121 \AA$^3$ and 
contained 20 and 10 atoms respectively. Ionic coordinates and cell parameters were relaxed at the DFT level until the residual forces on atoms were less than 0.02\,eV/\AA. PBEsol exchange–correlation functional \cite{perdew2008restoring} was used. A planewave cutoff of 700 eV was used which ensures the total energy to be converged to 1\,meV/atom. Standard PAW datasets supplied with VASP were employed \cite{blochl1994projector,kresse1999ultrasoft}. The valence configurations were Bi: 5$d^{10}$6$s^{2}$6$p^{3}$, Na: 2$s^{2}$2$p^{6}$3$s^{1}$, Ti: 3$s^{2}$3$p^{6}$3$d^{3}$4$s^{1}$, and O: 2$s^{2}$2$p^{4}$.

Starting from the relaxed structures and DFT wavefunctions, quasiparticle band gaps were obtained from single-shot $G_0W_0$ calculations as implemented in VASP \cite{shishkin2006implementation,shishkin2007self,shishkin2007accurate,fuchs2007quasiparticle}. %250 bands were used for self-energy calculation. %The energy cutoff for the response function was set to 250 eV. 80 grid points were used to represent Green’s function on the imaginary frequency axis. 
We used the GW-optimized PAW datasets distributed with VASP. The corresponding valence configurations were Bi: 5$s^{2}$5$d^{10}$6$s^{2}$6$p^{3}$, Na: 2$s^{2}$2$p^{6}$3$s^{1}$, Ti: 3$s^{2}$3$p^{6}$3$d^{4}$, and O: 2$s^{2}$2$p^{4}$. Workflow execution was automated with atomate2 \cite{ganose2025atomate2}.

\textbf{NaNbO$_3$}:
 For orthorhombic NaNbO$_3$ (\textit{Pbcm}), the experimental CIF \cite{mishchuk2004structural} was used as the starting structure (lattice parameters $a=5.50$\,\AA, $b=5.56$\,\AA, $c=15.54$\,\AA; $\alpha=\beta=\gamma=90\,^{\circ}$). The structure was relaxed at the DFT level using the same exchange--correlation functional, plane-wave cutoff, and force-convergence threshold as described above, with a Monkhorst--Pack $10\times10\times4$ $\mathbf{k}$-point mesh. A 40-atom primitive cell was employed. The optimized lattice constants are $a=5.49$\,\AA, $b=5.55$\,\AA, and $c=15.43$\,\AA, with $\alpha=\beta=\gamma=90\,^{\circ}$ which are in good agreement with experiment. The PAW pseudopotentials were employed for both the DFT geometry optimization and the $GW$ calculations, using Na $2s^2\,2p^6\,3s^1$, Nb $4s^2\,4p^6\,4d^4\,5s^1$, and O $2s^2\,2p^4$ valence configurations. Quasiparticle band gaps were obtained from single-shot $G_0W_0$ calculations using a $4\times4\times2$ $\mathbf{k}$-point mesh.

\section{Polaron conduction}
\label{polaron}

 Depending on the type of polarons present, electrical conduction can occur via small or large polaron transport \cite{Austin69, franchini21, Stone07, Shluger93, natanzon20}. In this work, we focus on small polaron conduction, which is considered the more likely mechanism in NBT and NN systems.

\begin{figure}[ht]
    \centering
    \includegraphics[width=10.5cm]{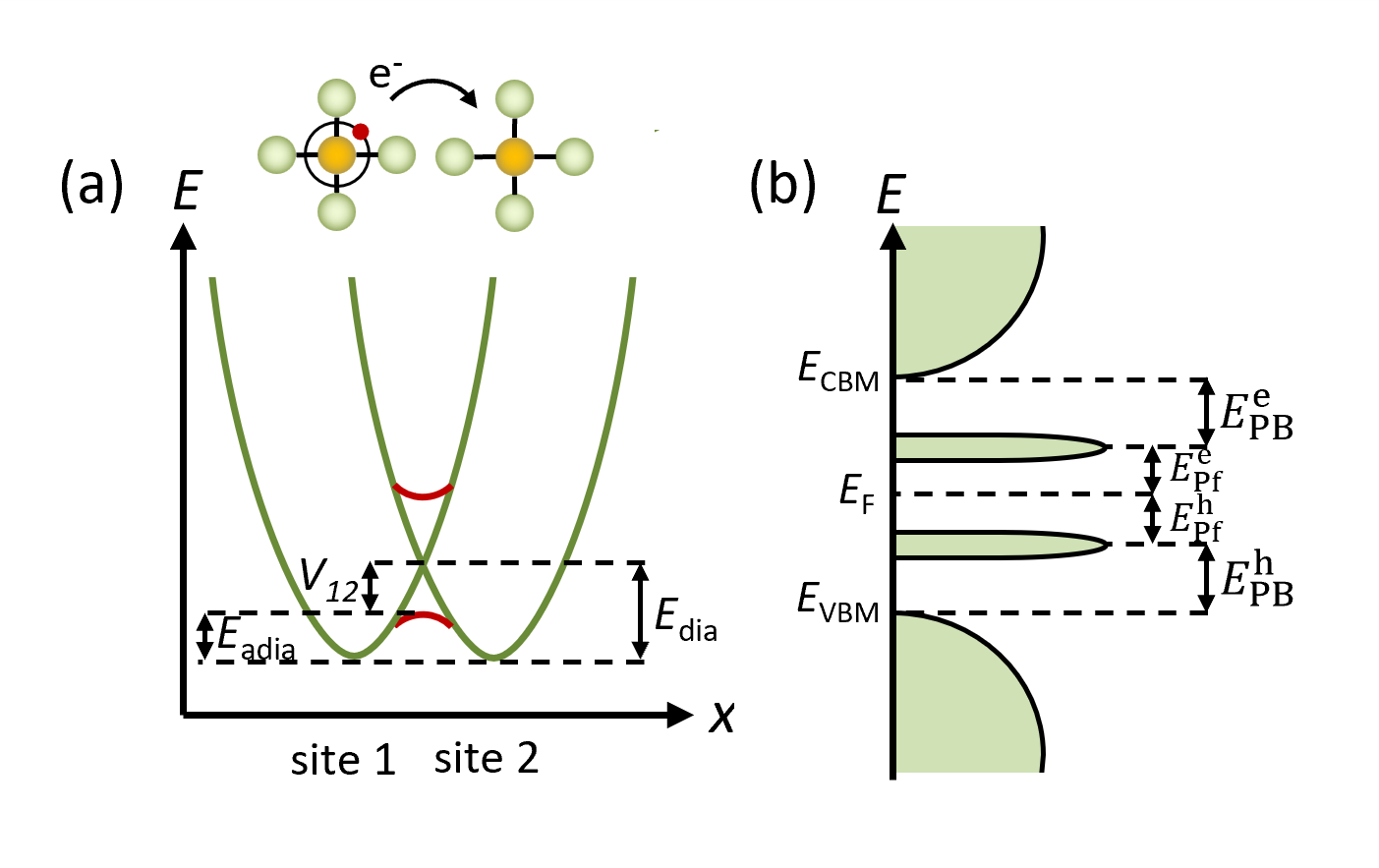}
    \caption{(a) Polaron hopping between two adjacent sites for the adiabatic and diabatic cases, characterized by activation energies $E_{\mathrm{adia}}$ and $E_{\mathrm{dia}}$, respectively  (adapted from \cite{Austin69}). $V_{12}$ denotes the electronic coupling strength. (b) Schematic band structure showing a hole and an electron polaron band, with polaron binding energy $E_{\mathrm{PB}}$ and polaron formation energy $E_{\mathrm{Pf}}$ (adapted from \cite{Wu03}).}
    \label{model}
\end{figure}

Small polaron conduction involves the thermally activated hopping of a localized charge carrier—either an electron or a hole—from one trapping site to an adjacent site. This process is illustrated in  Fig. \ref{model} (a), where two potential wells represent neighboring sites. For a carrier to hop from site 1 to site 2, the carrier must overcome the energy barrier separating the wells.

Polaron transport can occur in two distinct regimes: adiabatic and non-adiabatic (diabatic). In the adiabatic regime, typically relevant at low temperatures (e.g., when cooling from room temperature to 20\,$K$), the charge hopping frequency exceeds the phonon frequency. This allows for tunneling between sites due to significant orbital overlap, characterized by the electronic coupling strength $V_{12}$ (see Fig. \ref{model} (a)) In contrast, the diabatic regime, dominant at high temperatures (e.g., when heating from room temperature to several hundred degrees Celsius), features weaker orbital overlap due to phonon, induced lattice vibrations, reducing the tunneling probability. As a result, the activation energy required for hopping is effectively higher in the diabatic case than in the adiabatic case, with the difference roughly corresponding to $V_{12}$.

\begin{figure}[ht]
    \centering
    \includegraphics[width=7cm]{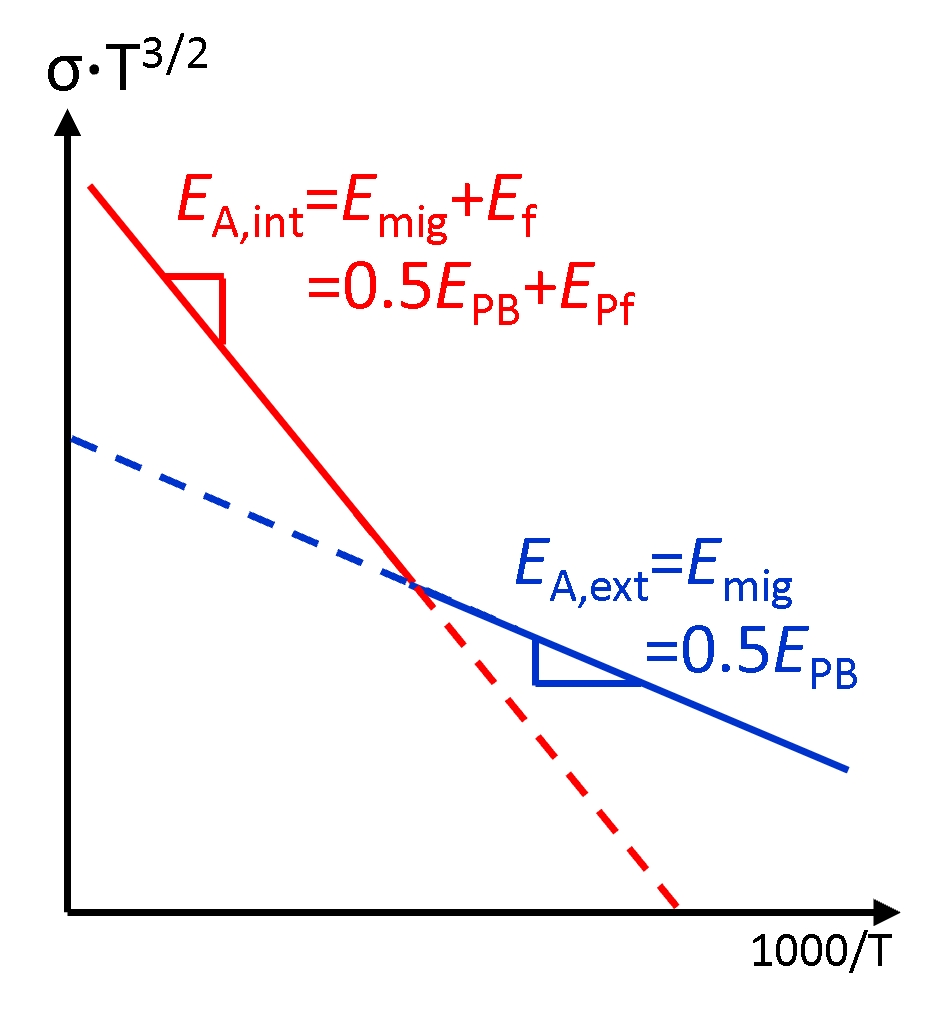}
    \caption{Identification of the extrinsic (blue) and ``intrinsic'' (red) regimes from the plot of $\sigma\cdot T^{3/2}$ as a function of 1000/T.}
    \label{regions}
\end{figure}

In general, the electrical conductivity due to polaron transport can be expressed as follows:
\begin{equation}
	\sigma=\frac{\sigma_{0}}{T^{A}}\exp\left (-\frac{E_{\mathrm{a}}}{k_{\mathrm{B}}T}\right )
\end{equation}
The prefactor $A$ equals 1 in the adiabatic regime and $\frac{3}{2}$ in the diabatic regime. The polaron concentration $c_{\mathrm{p}}$ and mobility $\mu_{\mathrm{p}}$ are given by Equations:
%
%\begin{equation}
%	\begin{array}{c}
%		c_{\mathrm{p}}=N_{\mathrm{p}}\exp\left (-\frac{E_{\mathrm{Pf}}}%{k_{\mathrm{B}}T}\right ) \\
%		\mu _{\mathrm{p}}=\frac{\mu_{0}}{T^{A}}\exp\left (-\frac{E_{\mathrm{(a)dia}}}{k_{\mathrm{B}}T}\right ) 
%	\end{array}
%\end{equation}
\begin{equation}
	\begin{aligned}
		c_{\mathrm{p}}=&N_{\mathrm{p}}\exp\left (-\frac{E_{\mathrm{Pf}}}{k_{\mathrm{B}}T}\right ) \\
		\mu _{\mathrm{p}}=&\frac{\mu_{0}}{T^{A}}\exp\left (-\frac{E_{\mathrm{(a)dia}}}{k_{\mathrm{B}}T}\right ) 
	\end{aligned}
\end{equation}
Here, $N_{\mathrm{p}}$ denotes the number of available polaron sites, and $E_{\mathrm{Pf}}$ represents the polaron formation energy, defined as the energetic difference between the polaron level and the Fermi level $E_{\mathrm{F}}$. Fig. \ref{regions} (b) illustrates the formation energies for both hole and electron polarons. Additionally, the polaron binding energy $E_{\mathrm{PB}}$ is indicated, which corresponds to the energy difference between the polaron level and the valence band maximum (VBM) for a hole polaron, or the conduction band minimum (CBM) for an electron polaron. The relationship between the polaron binding energy and the (a)diabatic activation energy is given by:
\begin{equation}
	\label{polaron equation}
	\begin{aligned}
		E_{\mathrm{PB}}=&2(E_{\mathrm{adia}}+V_{\mathrm{12}}) \quad \text{adiabatic regime} \\
		E_{\mathrm{PB}}=&2E_{\mathrm{adia}} \quad\quad\quad\quad\,\,\, \text{diabatic regime}		
	\end{aligned}
\end{equation}
Accordingly, the overall activation energy $E_{\mathrm{A}}$ for polaron conductivity can be expressed as:
\begin{equation}
	\begin{aligned}
		E_{\mathrm{A}}=&E_{\mathrm{Pf}}+E_{\mathrm{adia}}=E_{\mathrm{Pf}}+\frac{E_{\mathrm{PB}}}{2}-V_{\mathrm{12}} \quad \text{adiabatic regime} \\
		E_{\mathrm{A}}=&E_{\mathrm{Pf}}+E_{\mathrm{adia}}=E_{\mathrm{Pf}}+\frac{E_{\mathrm{PB}}}{2} \quad\quad\quad\,\,\,\, \text{diabatic regime}		
	\end{aligned}
\end{equation}

In this work, the analysis is restricted to the diabatic regime, which aligns with the employed conductivity measurements in the range between room temperature and $450 - 600\,\DEC$.

Typical Arrhenius plots are obtained using the product of the conductivity and T$^{3/2}$ as a function of 1000/T, as shown in Fig. \ref{regions}, which displays two regimes: extrinsic and ``intrinsic''. The extrinsic regime refers to a temperature range where the polaron concentration is fixed by external factors such as dopants or other defects. Under these conditions, the activation energy corresponds to half of the polaron binding energy:
\begin{equation}
    \begin{aligned}
        \sigma_{\mathrm{ext}}&=c_{\mathrm{p}}\cdot \mu_{\mathrm{p}}\cdot q   \\
        &=c_{\mathrm{p,ext}}\cdot\frac{\mu_{0}}{T^{3/2}}\exp\left(-\frac{E_{\mathrm{PB}}}{2k_{\mathrm{B} }T}\right)\cdot q   \\
        &=\frac{\sigma_{0}}{T^{3/2}}\exp\left(-\frac{E_{\mathrm{PB}}}{2k_{\mathrm{B}}T}\right)  
        \Rightarrow  E\mathrm{_{A,ext}}=\frac{E\mathrm{_{PB}^{e/h}}}{2}
        \end{aligned}
\end{equation}
The ``intrinsic'' regime, by contrast, occurs at higher temperatures where sufficient thermal energy is available to overcome the polaron (transport) gap, $E_{Pf}^e + E_{Pf}^h$, leading to dominant intrinsic polaron conduction. Here, the activation energy accounts for both the formation energy of the polarons and half of their binding energy:
\begin{equation}
	\begin{aligned}
       \sigma_{\mathrm{int}}&=c_{\mathrm{p}}\cdot \mu_{\mathrm{p}}\cdot q   \\
       &=N_{\mathrm{p}}\exp\left(-\frac{E_{\mathrm{Pf}}}{k_{\mathrm{B}}T}\right)\cdot\frac{\mu_{0}}{T^{3/2}}\exp\left(-\frac{E_{\mathrm{PB}}}{2k_{\mathrm{B} }T}\right)\cdot q \\
    &=\frac{\sigma_{0}}{T^{3/2}}\exp\left(-\frac{E_{\mathrm{Pf}}+\frac{E_{\mathrm{PB}}}{2}}{k_{\mathrm{B}}T}\right)	  \Rightarrow
		E\mathrm{_{A,int}}=\frac{E\mathrm{_{PB}^{e/h}}}{2}+E\mathrm{_{Pf}^{e/h}}
	\end{aligned}
\end{equation}

\section{Optical transitions} 
\label{optical}

\subsection{NBT-6BT}

The optical band gap was determined from diffuse reflectance (DR) measurements using Tauc’s equation \cite{elliott57, jubu20}.
\begin{equation}
    (\alpha h\nu )=C(h\nu-E_{\mathrm{g}})^{q} 
\end{equation}
where $\alpha$ is the absorption coefficient, $h$ is Planck's constant (4.135$\times$10$^{-15}$ eV$\cdot$s), $\nu$ is photon’s frequency (s$^{-1}$), $h\nu$ is incident photon energy, $C$ is a constant, $E_{\mathrm{g}}$ is the optical band gap, and $q$ is an exponent that depends on the nature of the electronic transition—with $q = 1/2$ for direct allowed transitions and $q = 2$ for indirect allowed transitions \cite{fox2010optical}. In a direct transition, the electron moves from the valence band to the conduction band without a change in crystal momentum, so photon absorption alone provides the required energy (as in GaAs or BaTiO$_3$). In contrast, an indirect transition requires a simultaneous phonon interaction to conserve momentum, as in Si or SrTiO$_3$ \cite{Yu01,pankove2012optical}.

For DR measurements, a Tauc plot was constructed by replacing the absorption coefficient $\alpha$ with the Kubelka–Munk function $F(R_\infty)$, which transforms the raw reflectance data into an equivalent absorption spectrum. The Kubelka–Munk function is defined as \cite{kubelka1931article,landi22}:
\begin{equation}
\alpha=F(R_{\infty})=\frac{K}{S}=\frac{(1-R_{\infty})^{2}}{2R_{\infty}}
\end{equation}
where $F(R_{\infty})$ is the K-M function, $K$ and $S$ are the absorption and scattering coefficients, respectively, and $R_{\infty}$ is the reflectance of an optically thick sample relative to that of a reference material \cite{jubu24}.

\begin{figure}[ht]
    \centering
    \includegraphics[width=9cm]{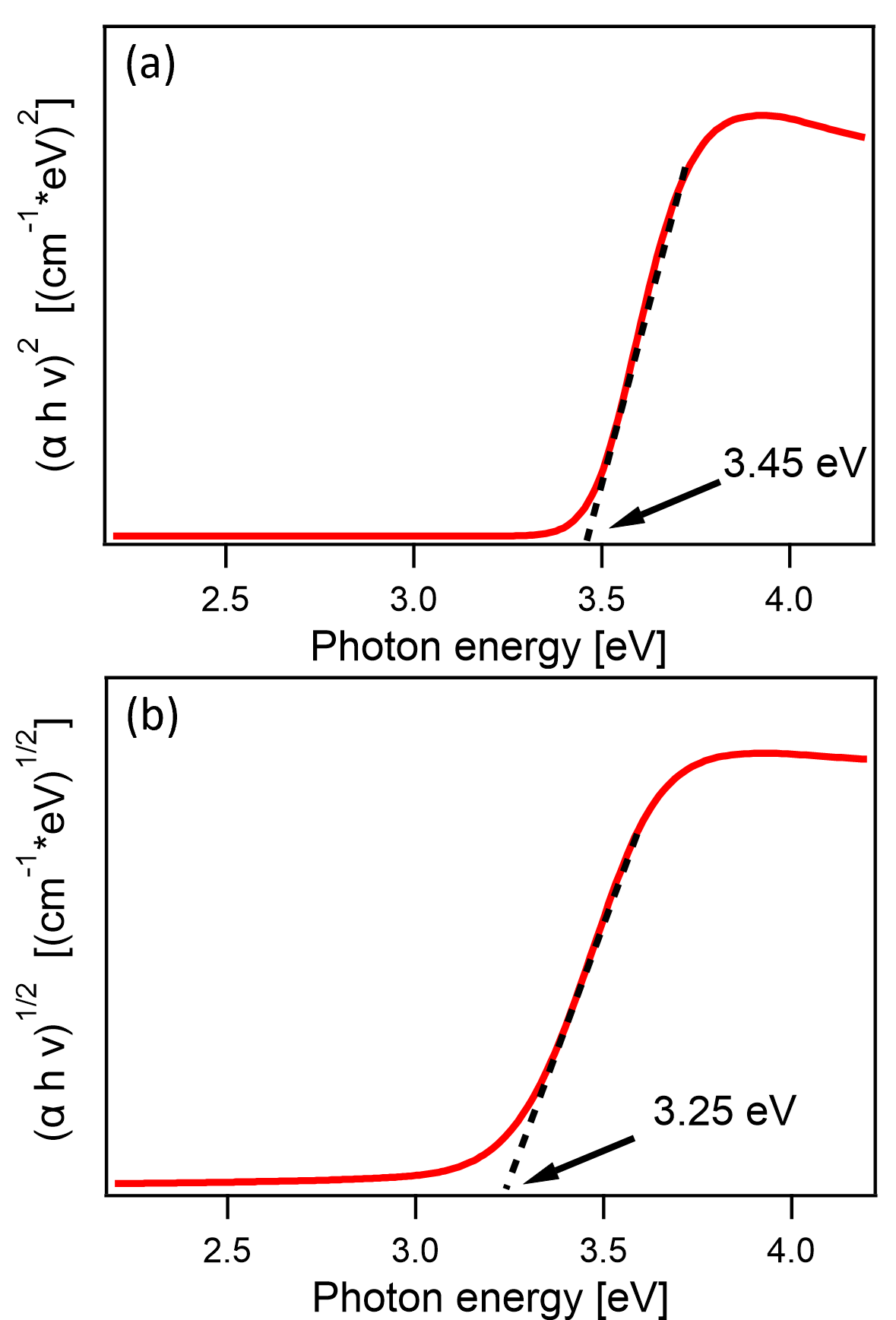}
    \caption{Tauc plots of NBT-6BT for (a) representation for a direct allowed transition: $(\alpha h\nu)^{2}=f(h\nu)$ and (b) for an indirect allowed transition: $(\alpha h\nu)^{1/2}=f(h\nu)$.}
    \label{NBT-Tauc}
\end{figure}

Fig. \ref{NBT-Tauc} (a) and (b) show the Tauc plots assuming direct and indirect allowed transitions, respectively, where $(\alpha h\nu)^{2}$ and $(\alpha h\nu)^{1/2}$ are plotted as a function of the incident photon energy. The optical band gaps, estimated from the intersection of the linear portion of the curve with the photon energy axis, are approximately 3.45 eV and 3.25 eV, respectively.

\subsection{NaNbO$_3$}

Bein et al.\ employed electron energy loss spectroscopy (EELS) in a transmission electron microscope to determine the optical band gap of NaNbO$_3$, obtaining a value of approximately 3.53\,eV \cite{bein22}. This result lies within the $3.4 - 3.5\,$eV range reported by optical spectroscopy in literature \cite{khorrami2015structural,kato02,zhu06,li2010band}.
%Gh. H. Khorrami previously reported an optical band gap of 3.40 eV for NaNbO$_3$ ceramics, also determined by the Kubelka–Munk method using the Tauc relation from UV–Vis diffuse reflectance spectroscopy \cite{khorrami2015structural}. This value falls within the range of 3.4 – 3.5 eV, as determined by optical spectroscopy in other studies \cite{khorrami2015structural,kato02,zhu06,li2010band}.

\section{X-ray photoelectron spectroscopy} 
\label{XPS}

The X-ray photoelectron spectroscopy (XPS) analysis used in this work have been already been reported elsewhere \cite{hu24,bein22}.

\subsection{NBT-6BT } 

The electron-trapped level in NBT–6BT was determined using in-situ XPS. Polycrystalline bulk ceramic samples were coated with 2-3\,nm thick layers of 10\% Sn-doped In$_2$O$_3$ (ITO). When a positive voltage was applied from the bottom Pt electrode across the cell, cathodic polarization of the ITO electrode attracted oxygen vacancies, leading to a reduction of the ITO. Metallic Bi was detected when the Fermi energy at the NBT–6BT surface increased to $E\mathrm{_F} - E\mathrm{_{VB}} = 2.47 \pm 0.10\,$eV, indicating that the electron-trapped level is associated with the Bi$^{3+/0}$ charge transition level and corresponds to $E\mathrm{_F} – E\mathrm{_{VB}} = 2.47 \pm 0.10\,$eV (as shown in Fig. \ref{xps-results}(a)) \cite{hu24}.

\subsection{NaNbO$_3$ }
The band gap of NaNbO$_3$ should be larger than the 3.4–3.5\,eV range reported in the literature based on optical spectroscopy and electron energy loss spectroscopy \cite{bein22}. This conclusion is supported by in situ XPS and electrical conductivity measurements. Specifically, the valence band maximum of NaNbO$_3$ was determined from Schottky barrier heights at interfaces with low–work function Sn-doped In$_2$O$_3$ and high–work function RuO$_2$, using XPS with in situ interface preparation. The results indicate that the valence-band edge of NaNbO$_3$ is comparable to those of SrTiO$_3$ and BaTiO$_3$. Considering that SrTiO$_3$ and BaTiO$_3$ have band gaps of approximately 3.2\,eV, close to the 3.4–3.5\,eV values previously reported for NaNbO$_3$, the conduction band minimum of NaNbO$_3$ is expected to be at a similar energy level. Based on this band alignment, donor doping would be expected to produce electrical conductivities on the order of 1\,S/cm, similar to donor-doped SrTiO$_3$ and BaTiO$_3$. In contrast, bulk ceramics of Sr- and Ca-doped NaNbO$_3$ exhibit room-temperature conductivities of only 10$^{-10}\,$S/cm, barely higher than undoped NaNbO$_3$. Moreover, high-field conductivity and impedance spectroscopy show no evidence that this extremely low conductivity arises from insulating grain boundaries. Therefore, the band gap of NaNbO$_3$ should exceed 3.4–3.5\,eV.

\begin{figure}[ht]
    \centering
    \includegraphics[width=0.7\linewidth]{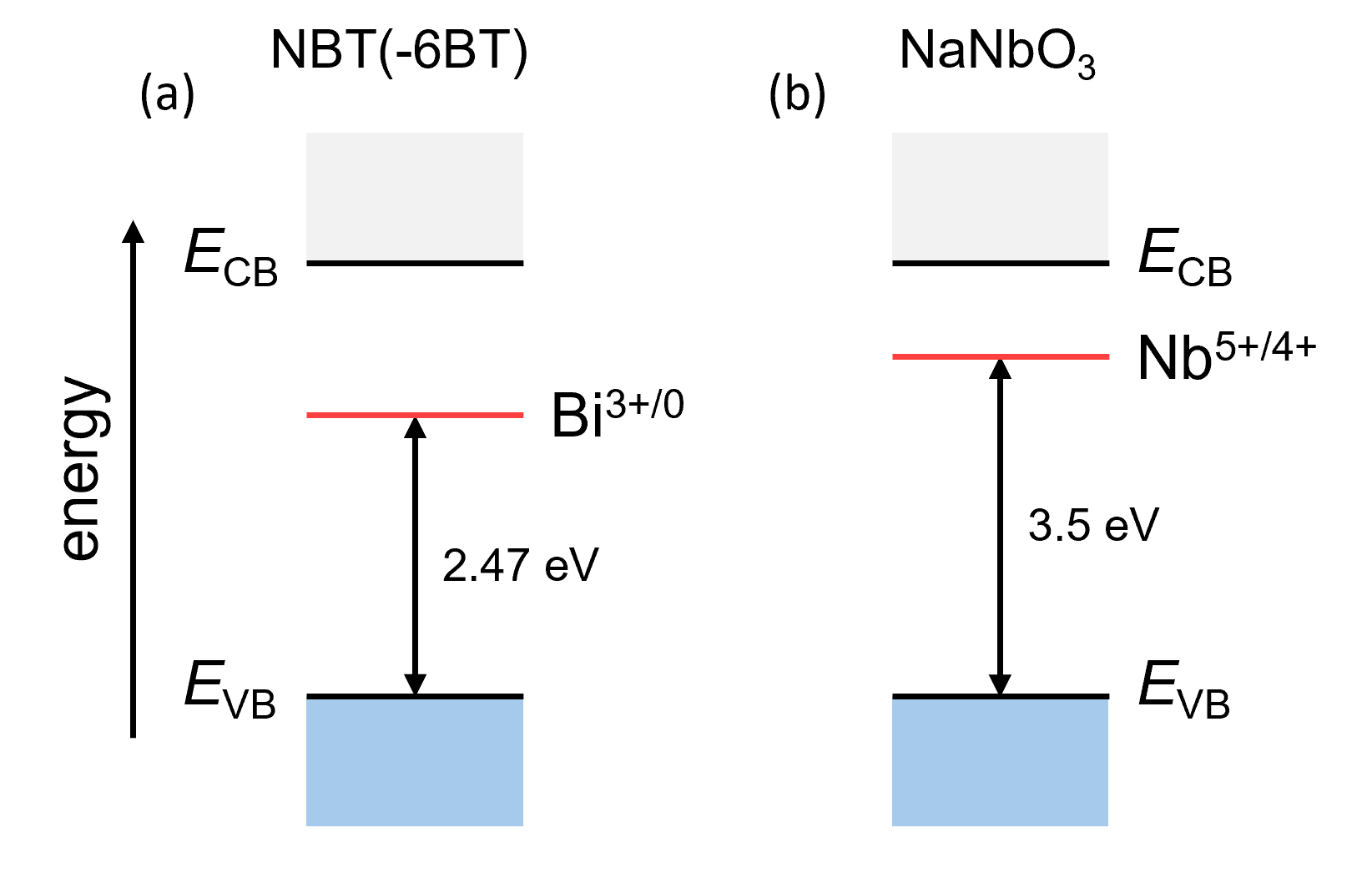}
    \caption{The upper limit of the Fermi level, corresponding to the energy distance between the electron trap level and the valence band maximum (VBM) in NBT-6BT \cite{hu24} and NaNbO$_3$ \cite{bein22}. The electron trap sites are associated with the Bi$^{3+/0}$ in NBT-6BT \cite{hu24} and the Nb$^{5+/4+}$ in NaNbO$_3$ \cite{schirmer1978two, sweeney1983oxygen, conradi2008influence}. }
    \label{xps-results}
\end{figure}

Moreover, the upper limit of the Fermi level, corresponding to the energy distance between the electron trap level and the VBM, was also determined to be 3.5\,eV from the interface experiment between ITO and NaNbO$_3$ (as shown in Fig. \ref{xps-results}(b)). However, the specific trapping site could not be identified; according to the literature, it is presumed to be associated with Nb sites involving the Nb$^{5+/4+}$ in this work \cite{schirmer1978two,sweeney1983oxygen,conradi2008influence}. In addition, hole polarons are also primarily assumed to be located at oxygen sites in this work \cite{schirmer1978two,conradi2008influence,choi1986electronic}.

\section{Electrical transport measurements}
\label{electrical}

\subsection{NBT-6BT}

Fig. \ref{sigmaT}(a) and (b) show the product of the conductivity and T$^{3/2}$ plotted as a function of 1000/T for Na$_{0.51}$Bi$_{0.49}$TiO$_3$-6BT (51/49) and Na$_{0.50}$Bi$_{0.50}$TiO$_3$-6BT (50/50) samples. As seen in Fig. \ref{sigmaT}(a), the 51/49 sample exhibits p-type conduction across the whole temperature region, evidenced by its higher conductivity in dry air compared to N$_2$. An extrinsic region is observed at lower temperatures, with an activation energy ($E_a$) of 0.54\,eV for both atmospheres, while the ``intrinsic'' region at higher temperatures shows E$_a$ values of 1.10\,eV in dry air and 1.11\,eV in N$_2$.

In contrast, the 50/50 sample (Fig. \ref{sigmaT}(b)) shows no clear extrinsic regime but only an ``intrinsic'' region, making it difficult to determine the binding energy of trapped charges. However, a transition in the dominant carrier type is observed within this ``intrinsic'' region near 320$\,^{\circ}$C, shifting from hole to electron polaron conduction, with $E_a$ values of 1.39\,eV in dry air and 1.56\,eV in N$_2$. The sample showing the transition between n- and p-type conduction and the two different activation energies will be utilized later to determine the transport gap.

\subsection{NaNbO$_3$}

\begin{figure}[ht]
    \centering
    \includegraphics[width=13.5cm]{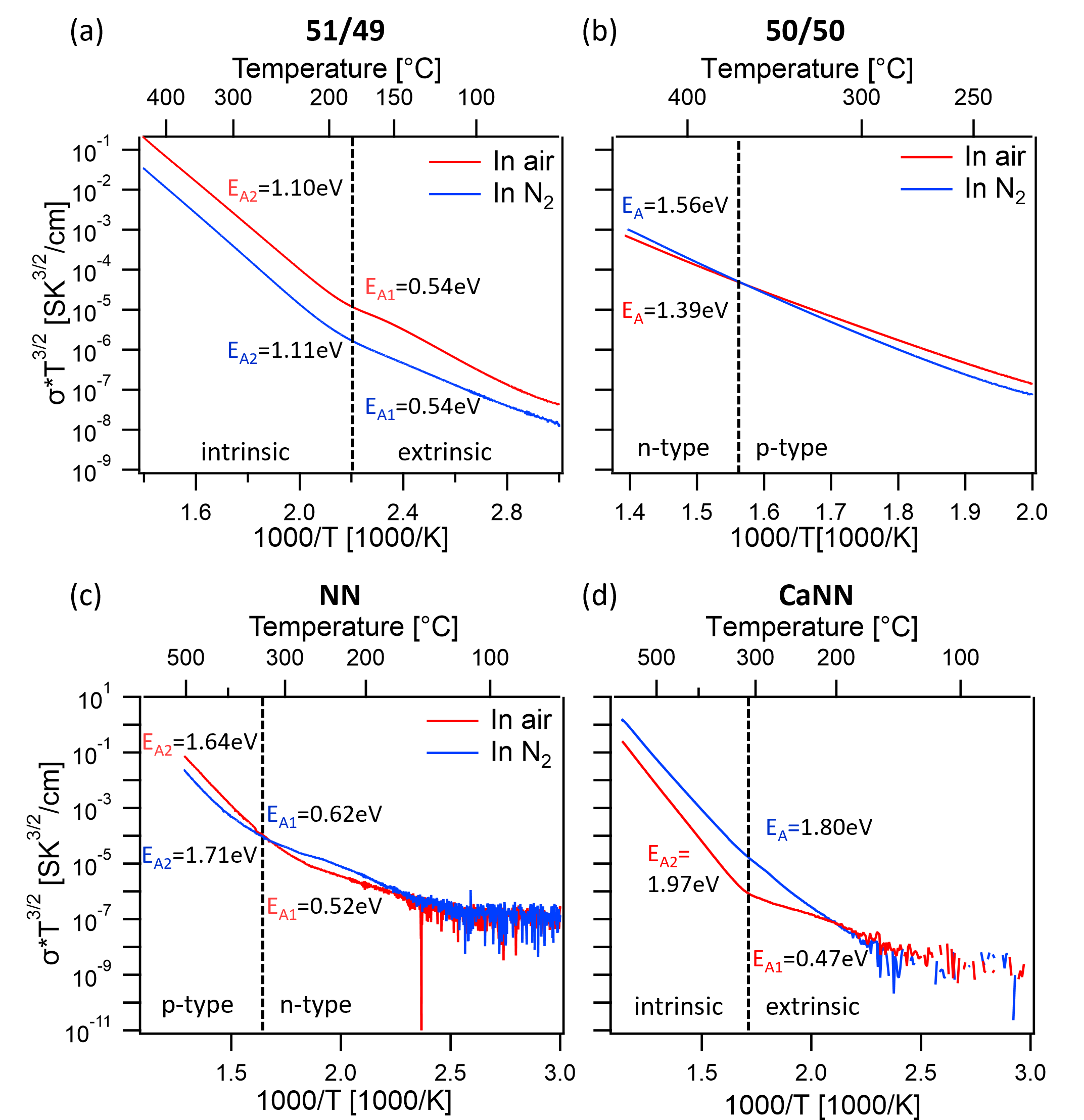}
    \caption{The product of the DC conductivity and T$^{3/2}$ as a function of 1000/T for (a) Na$_{0.51}$Bi$_{0.49}$TiO$_3$-6BT (51/49), (b) Na$_{0.50}$Bi$_{0.50}$TiO$_3$-6BT (50/50), (c) NaNbO$_3$ (NN) and (d) Ca-doped NaNbO$_3$ (CaNN).}
    \label{sigmaT}
\end{figure}

Fig. \ref{sigmaT}(c) and (d) shows the product of conductivity and T$^{3/2}$ plotted as a function of 1000/T for NN and CaNN samples. The NN sample exhibits n-type conduction in the extrinsic region at lower temperatures and p-type conduction in the ``intrinsic'' region at higher temperatures. In contrast, the Ca-doped sample exhibits n-type conduction across the entire temperature range, with a pronounced extrinsic region in dry air. The average activation energy ($E_a$) in the extrinsic region for both samples is approximately 0.5\,eV. 

%% following commands are inserted for easier reading 
\clearpage
\newpage

\section{Polaron binding and formation energies}
\label{polaronenergy}

The fundamental and transport gaps of NBT-6BT and NaNbO$_3$ ceramics are now determined based on polaron conduction behavior observed in dynamic, temperature-dependent measurements combined with XPS analysis. To derive the gaps, the electron and hole polaron binding energies and formation energies (see Fig.\ref{model}) have to be extracted from the measurements.

\subsection{NBT-6BT}

The binding energy of trapped holes ($E^{\mathrm{h}}_{\mathrm{PB}}$) is twice the activation energy in the extrinsic region ($E^{\mathrm{p}}_{\mathrm{a,ext}}$) of the p-type 51/49 sample (Fig. \ref{sigmaT}(a)). It is derived as 1.08\,eV. 

In contrast, the binding energy of trapped electrons ($E^{\mathrm{e}}_{\mathrm{PB}}$) is determined indirectly from the n-type 50/50 sample (Fig. \ref{sigmaT}(b)). In this sample, a transition in the dominant carrier type is observed near 320$\,^{\circ}$C, shifting from hole to electron polaron conduction. From the p-type regime, the Fermi level position ($E^{\mathrm{FL}}_{\mathrm{VBM}}$) can be estimated using the formation energy of trapped holes ($E^{\mathrm{h}}_{\mathrm{Pf}}$):

\begin{equation}
\begin{aligned}
    E^{\mathrm{h}}_{\mathrm{Pf}}&=E^{\mathrm{p}}_{\mathrm{a,int}}-E^{\mathrm{h}}_{\mathrm{PB}}  
    =1.39 -\frac{1.08}{2}=0.85 \,\mathrm{eV} \\
    \bigtriangleup E^{\mathrm{FL}}_{\mathrm{VBM}}&=E^{\mathrm{h}}_{\mathrm{Pf}}+E^{\mathrm{h}}_{\mathrm{PB}}  
    =0.85 +1.08=1.93 \,\mathrm{eV}
\end{aligned}    
\end{equation}
Given that the energy difference between the Bi$^{3+/0}$ trap level and VBM is 2.47\,eV from the XPS measurement \cite{hu24}, the formation energy of trapped electrons ($E^{\mathrm{e}}_{\mathrm{Pf}}$) can be calculated as:
\begin{equation}
    \begin{aligned}
      E^{\mathrm{e}}_{\mathrm{Pf}}&=\bigtriangleup E^{\mathrm{Bi^{3+/0}}}_{\mathrm{VBM}}-E^{\mathrm{FL}}_{\mathrm{VBM}}=2.47 -1.93=0.54 \,\mathrm{eV}  
    \end{aligned}
\end{equation}
The binding energy for electron traps ($E^{\mathrm{e}}_{\mathrm{PB}}$) in the n-type ``intrinsic'' region is then:
\begin{equation}
    E^{\mathrm{e}}_{\mathrm{PB}}=2 \times (E^{\mathrm{e}}_{\mathrm{a,int}}-E^{\mathrm{e}}_{\mathrm{Pf}})=2\times(1.56-0.54)=2.04 \,\mathrm{eV}
\end{equation}

\subsection{NaNbO$_3$}

%This section are based on the dissertation of N. Bein, to which readers are referred for further details \cite{bein24}.

\begin{figure}[ht]
    \centering
    \includegraphics[width=13.5cm]{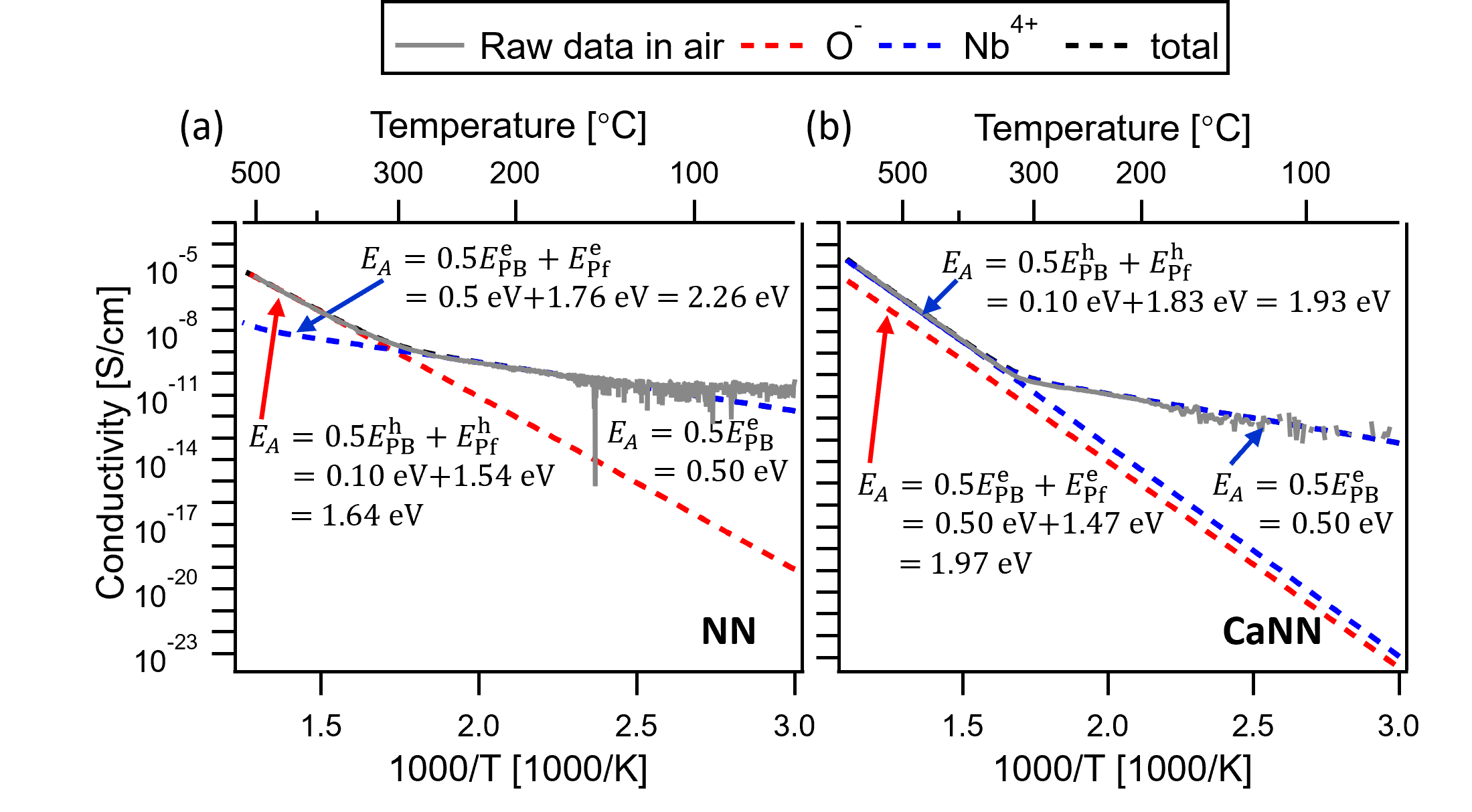}
    \caption{Measured and fitted conductivity of (a) NaNbO$_3$ and (b) Ca-doepd NaNbO$_3$ in air including all relevant energies.}
    \label{sigmaNN}
\end{figure}

Both NN and CaNN samples show n-type behavior in the extrinsic region (Fig. \ref{sigmaT}). The $E^{\mathrm{e}}_{\mathrm{PB}}$ is taken to be twice the average activation energy in the extrinsic region ($E^{\mathrm{n}}_{\mathrm{a,ext}}$), giving a value of 1.0\,eV (Fig. \ref{sigmaNN}(b)). We then apply the equations derived in Section \ref{polaron} to simulate the conductivity, considering that electrons are trapped at Nb sites (Nb$^{5+/4+}$) and holes at O sites (O$^{2-/-}$):
\begin{equation}
\begin{aligned}
   \text{NN:} \quad\sigma_{\mathrm{ext}}&=c_{\mathrm{Nb^{4+}}}\cdot\mu_{\mathrm{Nb^{4+}}}\cdot q \\
                &=[\mathrm{Nb^{'}_{Nb}}]_{\mathrm{ext}}\cdot\frac{\mu_{\mathrm{Nb^{4+},0}}}{T^{3/2}}\exp\left(\frac{E_{\mathrm{PB}}^{\mathrm{e}}/2}{k_{\mathrm{B}}T}\right)\cdot q  \\
                \sigma_{\mathrm{int}}&=c_{\mathrm{O^{-}}}\cdot\mu_{\mathrm{O^{-}}}\cdot q \\
                &=N_{\mathrm{O}}\exp\left(-\frac{E_{\mathrm{Pf}}^{\mathrm{h}}}{k_{\mathrm{B}}T}\right)\cdot\frac{\mu_{\mathrm{O^{-},0}}}{T^{3/2}}\exp\left(\frac{E_{\mathrm{PB}}^{\mathrm{h}}/2}{k_{\mathrm{B}}T}\right)\cdot q  \\     
\text{CaNN:} \quad  \sigma_{\mathrm{ext}}&=c_{\mathrm{Nb^{4+}}}\cdot\mu_{\mathrm{Nb^{4+}}}\cdot q \\
                &=[\mathrm{Nb^{'}_{Nb}}]_{\mathrm{ext}}\cdot\frac{\mu_{\mathrm{Nb^{4+},0}}}{T^{3/2}}\exp\left(\frac{E_{\mathrm{PB}}^{\mathrm{e}}/2}{k_{\mathrm{B}}T}\right)\cdot q  \\
                \sigma_{\mathrm{int}}&=c_{\mathrm{Nb^{4+}}}\cdot\mu_{\mathrm{Nb^{4+}}}\cdot q \\
                &=N_{\mathrm{Nb}}\exp\left(-\frac{E_{\mathrm{Pf}}^{\mathrm{e}}}{k_{\mathrm{B}}T}\right)\cdot\frac{\mu_{\mathrm{Nb^{4+},0}}}{T^{3/2}}\exp\left(\frac{E_{\mathrm{PB}}^{\mathrm{e}}/2}{k_{\mathrm{B}}T}\right)\cdot q  \\        
\end{aligned}
\end{equation}

Fig. \ref{sigmaNN} presents the measured and fitted conductivities of (a) NaNbO$_3$ and (b) Ca-doped NaNbO$_3$ in air, including all relevant energy parameters. The binding energy of trapped holes ($E^{\mathrm{h}}_{\mathrm{PB}}$) for CaNN was determined from the simulation parameters, as shown in Table \ref{The simulation parameter1}, to be no greater than 0.2\,eV when fulfilling the condition that the binding and formation energies of hole polarons are adjusted such that the ``intrinsic'' hole conductivity remains lower than the ``intrinsic'' electron conductivity.

Subsequently, $E^{\mathrm{e}}_{\mathrm{Pf}}$ and $E^{\mathrm{h}}_{\mathrm{Pf}}$ for NN (Fig. \ref{sigmaNN}(a)) were determined from the simulation parameters, also summarized in Table \ref{The simulation parameter1}, to be 1.76\,eV and 1.54\,eV, respectively.

\begin{table}[ht]
	\renewcommand\arraystretch{1.3}
	\centering
	\caption{Hole and electron polaron concentration and mobility at 500$\DEC$ in air for NaNbO$_3$ and Ca-doped NaNbO$_3$.}
	\label{The simulation parameter1}
	\vspace{7pt}
	\small
	\setlength{\tabcolsep}{2.5mm}{
		\begin{tabular}{lcccc}
			\hline \hline
			& $c_{\mathrm{O}^{-}}$ [cm$^{-2}$]& $\mu_{\mathrm{O}^{-}}$ [cm$^2$V$^{-1}$s$^{-1}$]& $c_{\mathrm{Nb}^{4+}}$ [cm$^{-2}$]& $\mu_{\mathrm{Nb}^{4+}}$ [cm$^2$V$^{-1}$s$^{-1}$]\\ \hline
			NN   &4.24$\cdot$10$^{12}$  & 4.15  & 2.02$\cdot$10$^{11}$   & 0.42     \\
			CaNN &5.42$\cdot$10$^{10}$  & 4.15  & 4.05$\cdot$10$^{12}$  & 0.42  \\
			\hline \hline			
	\end{tabular}}
\end{table}

\section{The transport and fundamental gaps}
\label{TFgap}

\begin{figure}[ht]
    \centering
    \includegraphics[width=10.5cm]{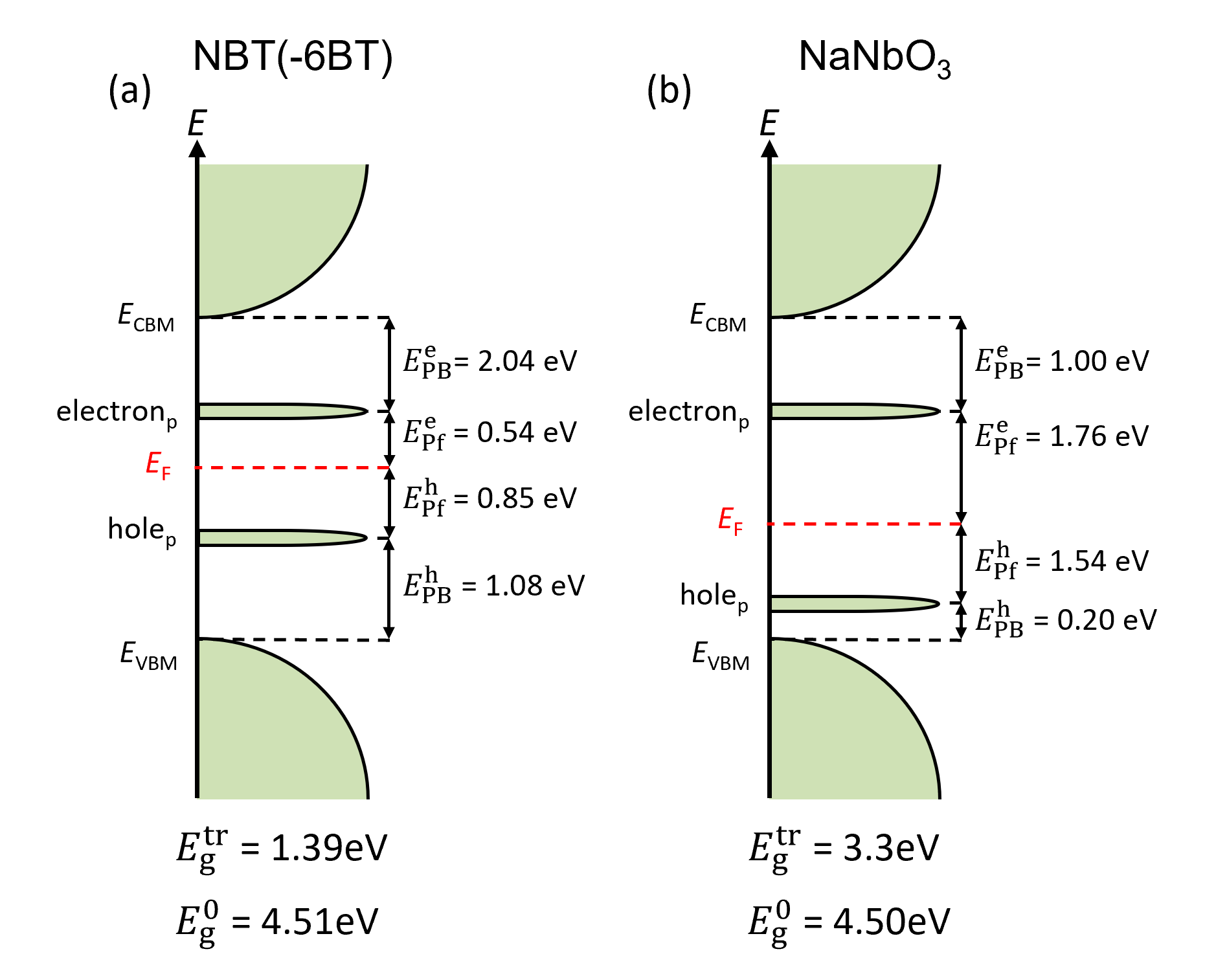}
    \caption{Schematic diagrams of the electronic band structures of (a) NBT-6BT and (b) NaNbO$_3$ determined by electrical methods. In both cases, the binding energies and formation energies of trapped electrons and holes are indicated, along with the position of the Fermi level. The fundamental band gaps ($E_{\mathrm{g}}^{\mathrm{0}}$) of 4.51\,eV and 4.50\,eV are the sums of both polaron's binding and formation energies.}
    \label{diagrams}
\end{figure}

Table \ref{The polaron energy} summarizes the polaron binding and formation energies for NBT-6BT and NN.  Based on these values, the transport band gaps ($E_\mathrm{g}^\mathrm{tr}$) for NBT-6BT and NN are determined from the sums of the electron and hole polaron formation energies (Fig.~\ref{model}) as 1.39\,eV and 3.3\,eV, respectively. The fundamental gaps are the sums of the electron and hole polaron formation and binding energies and are 4.51\,eV for NBT-6BT and 4.5\,eV for NaNbO$_3$. The electronic band structures determined from these numbers are depicted in Figure \ref{diagrams}.

\begin{table}[ht]
	\renewcommand\arraystretch{1.3}
	\centering
	\caption{Polaron binding ($E_\mathrm{PB}$) and formation ($E_\mathrm{Pf}$) energies of electrons  holes derived from experiment. The transport gaps ($E_\mathrm{g}^\mathrm{tr}$) are the sums of the polaron formation energies and the fundamental gaps ($E_\mathrm{g}^\mathrm{0}$) are the sums of the binding and formation energies of trapped electron and holes.}
	\label{The polaron energy}
	\vspace{7pt}
	\small
	\setlength{\tabcolsep}{2.5mm}{
		\begin{tabular}{lcccccc}
			\hline \hline
			& $E\mathrm{_{PB}^{e}}$ [eV]& $E\mathrm{_{Pf}^{e}}$ [eV]& $E\mathrm{_{PB}^{h}}$ [eV]& $E\mathrm{_{Pf}^{h}}$  [eV] &$E_\mathrm{g}^\mathrm{tr}$ [eV]& $E\mathrm{_{g}^{0}}$ [eV]\\ \hline
			NBT-6BT   & 2.04  & 0.54   & 1.08   & 1.795  &1.39  &  4.51  \\
			NN        & 1.0   & 1.76    & 0.2  & 1.54  &3.3 &  4.5\\
			\hline \hline			
	\end{tabular}}
\end{table}

%In NBT-6BT, electrons are trapped at Bi sites (Bi$^{3+/0}$), as determined by XPS in Section~\ref{XPS}, while holes may be trapped either at Bi sites (Bi$^{3+/4+}$) or at O sites (O$^{2-/-}$). For NN, it is assumed that electrons are trapped at Nb sites (Nb$^{5+/4+}$) and holes at O sites (O$^{2-/-}$).

\section{DFT calculation of the fundamental gaps}
\label{DFT}

\begin{figure}[ht]
    \centering
    \includegraphics[width=13.5cm]{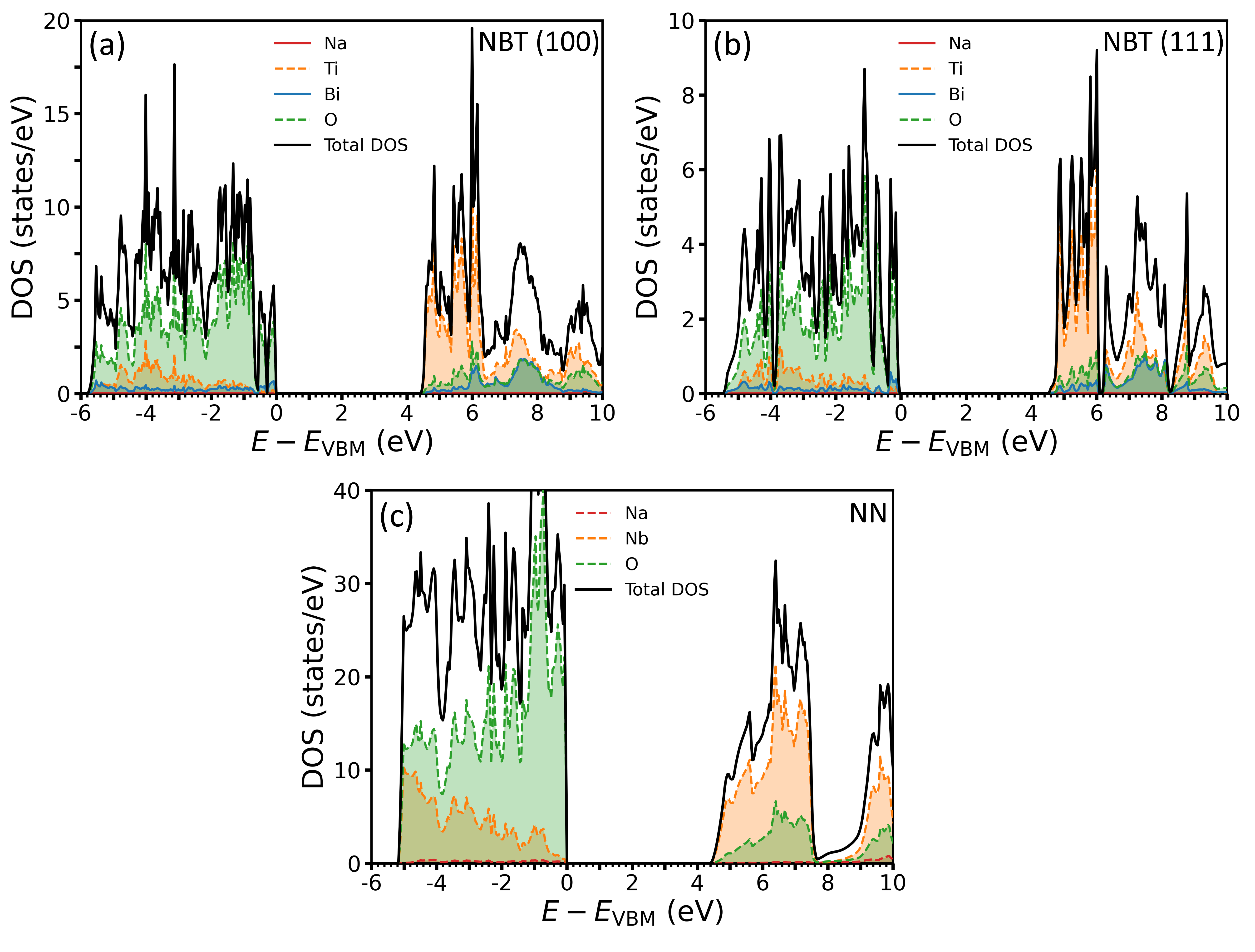}
    \caption{The projected density of states (pDOS) for cubic Na$_{0.5}$Bi$_{0.5}$TiO$_3$ with (a) (100) and (b) (111) $A$-site cation orderings and (c) NaNbO$_3$ calculated using density functional theory within the GW approximation. The different colors indicate the major orbital contributions to the electronic states.}
    \label{DOS}
\end{figure}

In order to validate the experimentally derived fundamental gaps, GW calculations are performed. Fig. \ref{DOS}(a) and (b) depict the projected density of states (pDOS) for cubic Na$_{0.50}$Bi$_{0.50}$TiO$_3$ (NBT) with $A$-site (100) and (111) cation orderings. The two orderings have different total energies  \cite{groeting11} but yield very similar density of states. In both cases the upper valence band is dominated by O $2p$ states that hybridize with Bi $6s$; the Bi $6s$ weight at the very VBM is small, consistent with a lone-pair contribution deeper in the valence manifold. Bi $6p$ states appear in the lower valence region and are prominent throughout the conduction band. The conduction-band minimum is primarily Ti $3d$ in character with an appreciable admixture of Bi $6p$, while Na states contribute negligibly near the band edges. As for the fundamental band gaps, both cases show values of approximately 4.5\,eV, in good agreement with the experimentally derived value.

For comparison, Fig. \ref{DOS}(c) presents the pDOS of orthorhombic NaNbO$_3$ computed with the same methodology. The valence band is largely O $2p$ in character with minor Nb $4d$ hybridization, and the conduction-band onset is dominated by Nb $4d$ states. Na contributions are weak away from the band edges. The fundamental band gap is approximately 4.40\,eV, which is also in good agreement with the experimentally derived value. The fundamental gap of NaNbO$_3$ of $4.4-4.5\,$eV and its difference to the optical gap of $3.4-3.5\,$eV are also comparable to those of LiNbO$_3$ \cite{schmidt08} and KNbO$_3$ \cite{schmidt19}.

\section{The semiconducting nature of ferroelectrics} 
\label{semifer}

The above results demonstrate that it is important to distinguish between different types of energy gaps when analyzing ferroelectric oxides. In organic semiconductors, the typical relationship follows: $E_{\rm g}^{0} > E_{\rm g}^{\rm tr} > E_{\rm g}^{\rm opt}$. In contrast, for the ferroelectric oxides studied here, the relationship is altered:  $E_{\rm g}^{0} > E_{\rm g}^{\rm opt} > E_{\rm g}^{\rm tr}$ for NBT-6BT and NN, indicating that the optical gap lies above the transport gap but below the fundamental gap.

It is worth noting that the concept of electron and hole traps provides greater flexibility and a more comprehensive framework for the study of defect chemistry. Trapping and detrapping of charge carriers enable the establishment of self-consistent correlations between conduction mechanisms and quantities such as electrical conductivity, activation energy, defect concentration, and carrier mobility, as discussed in Section \ref{polaronenergy}. This framework is applicable even to comparatively insulating materials, since, as emphasized by Fridkin \cite{fridkin}, the concentrations of trapped carriers are often several orders of magnitude higher than those of free carriers. On the other hand, the presence of electrons and holes in ferroelectrics is frequently described within the framework of conventional semiconductors, where charges occupy the valence and conduction bands. For instance, the quantitative modeling of charged domain-wall screening by mobile electrons and holes, which is responsible for the enhanced electronic conduction observed at domain walls \cite{seidel09}, adopts a classical semiconductor approach, assuming mobile carriers reside within the energy bands \cite{sluka12,zuo14}.

In ferroelectric oxides, the concentration of free electronic carriers is typically much smaller than that of ionic species ($[\rm e_{\mathrm{free}}] \ll [\mathrm{ion}]$). However, the concentration of trapped electrons can be comparable to, or even exceed, that of ionic defects ($[\rm e_{\mathrm{trap}}]\approx [\mathrm{ion}]$ or $[\rm e_{\mathrm{trap}}] > [\mathrm{ion}]$). This has important consequences for defect equilibria: for instance, the concentration of oxygen vacancies ($v_{\mathrm{O}}^{\cdot\cdot}$) after sintering is not necessarily determined by the classical Schottky equilibrium with cation vacancies but may instead be constrained by the trapping of electronic carriers, such that $2 [v_{\mathrm{O}}^{2+}] = [\rm e_{\mathrm{trap}}^-]$. Similarly, cation vacancies can be coupled to trapped holes, leading to $n \times [v_\mathrm{cat}^{n-}] = [\rm h_{\mathrm{trap}}^+]$ as dominant defect equilibrium. 

Regarding the energy gaps, Barium titanate (BaTiO$_3$) represents a notable exception among ferroelectric oxides, as it does not exhibit significant electron trapping in the absence of oxygen vacancies \cite{yoo2002batio3} and trapped holes have very low trapping energies \cite{erhart14, Traiwatt18, schirmer11}. For this material, there is also no indication of significant differences between the fundamental, optical and electrical energy gaps.

\section{Summary}

In summary, this work underscores the necessity of distinguishing between the fundamental, optical, and electrical energy gaps in ferroelectric oxides, as each reflects distinct physical processes and electronic states. The fundamental gap represents the intrinsic ground-state separation between the valence and conduction bands, whereas the optical and transport gaps are governed by excited-state phenomena involving localized polaronic carriers. These differences, typically on the order of 1 electronvolt, have significant implications for interpreting the electronic and optoelectronic behavior of ferroelectrics. By combining optical spectroscopy, X-ray photoelectron spectroscopy, and conductivity measurements under controlled temperature and oxygen partial pressure, this study provides a comprehensive framework for determining and rationalizing these distinct energy gaps in NBT-6BT and NaNbO$_3$, demonstrating the hierarchy $E{\rm g}^{0} > E_{\rm g}^{\rm opt} > E_{\rm g}^{\rm tr}$ in ferroelectric oxides. Additionally,  DFT calculation within the GW approximation confirm the plausibility of a  4.5\,eV fundamental band gaps in NBT and NaNbO$_3$. Such insights lay a solid foundation for tailoring ferroelectric compounds toward emerging applications in energy conversion and catalytic devices.

\ack
The presented work has been executed within the collaborative research centre FLAIR (Fermi level engineering applied to oxide electroceramics), which is funded by the German Research Foundation (DFG), project-ID 463184206 -- SFB 1548, and by the Austrian Fonds zur F{\"{o}}rderung der wissenschaftlichen Forschung (FWF),Project Grant-DOI (10.55776/I6450). Additional support was provided by the state of Hesse, Germany, within the LOEWE priority project FLAME (Fermi Level Engineering of Antiferroelectric Materials for Energy Storage and High Voltage Insulation Systems), by the German Academic Exchange Service (DAAD) through the PPP Slovenia, Project ID 57450108, and by the Slovenian Research Agency (core funding P2-0105 and bilateral Project No. BI-DE/19-20-008). Pengcheng Hu also acknowledges support from the China Scholarship Council (CSC), Award No. 202106220039. Paderborn supercomputing center is gratefully acknowledged by Mohammad Amirabbasi as the provider of needed computing facilities. Tadej Rojac and Barbara Mali\v{c} gratefully acknowledge the financial support of the Slovenian Research and Innovation Agency (research core funding P2-0105). For the purpose of open access, the authors have applied a CC BY public copyright license to any Author Accepted Manuscript version arising from this submission.

\section*{Data availability statement}

All data that support the findings of this study are included within the article (and any supplementary files).

%\appendix

\section*{References}

\bibliographystyle{iopart-num}

\bibliography{ferrogap}

\end{document}